\pdfoutput=1
\documentclass[aps,pra,reprint,superscriptaddress,amsmath,amssymb,longbibliography,floatfix]{revtex4-2}
\usepackage[T1]{fontenc}
\usepackage{lmodern}
\input{glyphtounicode}\pdfgentounicode=1
\usepackage[utf8]{inputenc}
\usepackage{graphicx}
\usepackage{bm}
\usepackage{booktabs}
\usepackage{amsmath}
\usepackage{amssymb}
\usepackage[expansion=false]{microtype}
\usepackage[colorlinks=true,linkcolor=blue,citecolor=blue,urlcolor=blue]{hyperref}

\graphicspath{{figures/}{./}}

\newcommand{\ket}[1]{\lvert #1\rangle}
\newcommand{\bra}[1]{\langle #1\rvert}
\newcommand{\Dstate}[2]{\ket{D_{#1}^{(#2)}}}
\newcommand{\sect}[2]{(#1,#2)}
\newcommand{\gamin}{\gamma_{\mathrm{in}}}
\newcommand{\gamsh}{\gamma_{\mathrm{sh}}}
\newcommand{\gam}{\boldsymbol{\gamma}}
\newcommand{\RZ}{R_Z}
\newcommand{\qfim}{\mathcal{F}}
\newcommand{\reff}{\mathrm{PR}}
\newcommand{\Loss}{\mathcal{L}}
\newcommand{\cproj}{\mathbb{CP}^{C-1}}
\newcommand{\SWAP}{\mathrm{SWAP}}
\DeclareMathOperator{\rank}{rank}
\DeclareMathOperator{\Lie}{Lie}
\DeclareMathOperator{\conv}{conv}
\DeclareMathOperator{\ord}{ord}

\begin{document}

\title{Differentiable collision-model control for multipartite state engineering: reachability, geometry, and preparation cost}

\author{Minh Tam Nguyen}
\affiliation{KT One, Tampa, Florida 33613, USA}

\author{Duc-Kha Vu}
\affiliation{Department of Electrical and Computer Engineering, Saint Louis University, St.~Louis, Missouri 63103, USA}

\author{\"Ozg\"ur E. M\"ustecapl{\i}o\u{g}lu}
\affiliation{Department of Physics, Ko\c{c} University, Istanbul, T\"urkiye}
\affiliation{T\"UB\.ITAK Research Institute for Fundamental Sciences (TBAE), Gebze, T\"urkiye}

\author{Fatih Ozaydin}
\email{mansursah@gmail.com}
\affiliation{Institute for International Strategy and Emerging Technologies, Tokyo International University, Tokyo 170-0013, Japan}
\affiliation{Data Science \& AI Major, Tokyo International University, Tokyo 170-0013, Japan}
\affiliation{Nanoelectronics Research Center, Istanbul, T\"urkiye}

\date{\today}

\begin{abstract}
We study a differentiable control architecture for multipartite state engineering in which partial-SWAP collisions, a Heisenberg exchange, are interleaved with layers of local phase rotations. Singular collision angles confine the evolution to invariant supports, impose exact fidelity ceilings, and generate an inverse-distance lower bound on the preparation time nearby. Away from them, numerical certificates show that the sampled couplings reach the whole fixed-excitation sector for sector dimensions up to 21, and optimizations reach fidelities above 0.999 for dimensions up to 1716. At generic couplings the number of rounds needed for a given Dicke fidelity, measured by the ratio of independent phase controls to the projective dimension, is nearly independent of the system size, and the quantum Fisher information spectrum is close to that of a random Gaussian Jacobian. Near full SWAP, a slow-time theory, the sharp form of the inverse-distance bound there, predicts microscopic hitting times without fitted parameters: the distance from the singular point sets the clock and its direction selects the effective control geometry. A detuning sweep along two rays shows that the Dicke horizon follows this law up to a detuning of 0.1, where it is two to five times shorter than at a generic coupling in sectors of dimension 70 to 252, while phase-uniform, random-sign, and Haar-random targets are delayed; beyond a detuning of about 0.3 all targets return to the generic cost. The slow-time value for the Dicke target is nearly the same for sector dimensions from 35 to 252. The exchange graph of the effective theory offers an interpretation of this ordering. Counted in two-qubit gates, generic-coupling preparation is costlier than known Dicke circuits; the architecture offers fixed global couplings, local phase rotations as the only control, and no measurement.
\end{abstract}

\maketitle

\section{Introduction}
\label{sec:intro}

Multipartite entangled states are resources for quantum sensing, communication, and computation, and preparing a prescribed state deterministically on finite hardware is a control problem. Dicke states provide a natural testbed. They are permutation-symmetric superpositions of all computational-basis states with a fixed number \(K\) of excitations among \(N\) qubits~\cite{dicke1954},
\begin{equation}
    \Dstate{N}{K}=\binom{N}{K}^{-1/2}
    \sum_{s\in\binom{[N]}{K}}\ket{s},
    \label{eq:dicke-def-intro}
\end{equation}
and the \(K=1\) member is the \(W\) state, whose entanglement survives the loss of particles~\cite{dur2001,neven2018}. Dicke states with more excitations have larger entanglement depth~\cite{chen2016} and remain useful for metrology under noise~\cite{saleem2024}. Their preparation has been approached with linear optics and fusion gates for \(W\) states~\cite{tashima2008,bugu2013,yesilyurt2013,ozaydin2014}, deterministic expansion with separable ancillas~\cite{yesilyurt2016,zang2016,ozaydin2021}, gate-based circuits with explicit gate counts~\cite{bartschi2019,chakraborty2014,mukherjee2020,aktar2022}, mid-circuit collective measurement with feedback~\cite{yu2024}, global or single-site control of coupled spins~\cite{stojanovic2023,stojanovic2026}, phase estimation in a spin ensemble~\cite{wangterhal2021}, adiabatic driving~\cite{linington2008,opatrny2016,carrasco2024}, engineered dissipation~\cite{zhu2025}, and transformations between Dicke states~\cite{kobayashi2014,thapa2025}. The adaptive protocol of Ref.~\cite{yu2024} reaches polylogarithmic resources with collective measurements and feedback, which shows that the cost of preparing a Dicke state is an open resource question even when the state is known to be reachable.

Collision models describe a system that interacts with a sequence of ancillary units one at a time~\cite{ciccarello2022,campbell2021}. They were developed for open-system dynamics and can simulate any multipartite Markovian evolution~\cite{cattaneo2021}, but the same repeated interactions can build entanglement. \c{C}akmak \textit{et al.} showed that a mobile shuttle qubit generates multipartite entanglement through excitation-preserving partial-SWAP collisions with register qubits that never interact directly~\cite{cakmak2019}. In Ref.~\cite{vu2026pra} we removed the shuttle measurement of that scheme: \(K\) initially excited shuttle qubits remain part of the final state, and the partial-SWAP sequence prepares multi-excitation Dicke states deterministically in principle, with only two global collision angles as parameters. Reference~\cite{vu2026data} translated the resulting depth--fidelity frontiers into circuit costs and ran selected circuits on a superconducting processor, and in a recent work~\cite{vu2026sd} we compared learned and classical strategies for choosing the two collision angles under a limited evaluation budget.

The present paper asks a different question about the same architecture. We hold the two collision angles fixed and insert a layer of local \(\RZ\) rotations after every collision round. The collisions mix amplitudes within the fixed-excitation sector, and the phase layers steer the interference among them. Over \(T\) rounds the control sequence has \(T(N-1)\) independent parameters after an exact gauge direction is removed from each layer, and the final state is a differentiable function of all of them. This raises three questions: which states of the sector are reachable at a given pair of collision angles, how well conditioned the local control geometry is, and how many rounds a prescribed target requires. Closely related questions arise for parametrized quantum circuits in variational algorithms~\cite{cerezo2021}, where expressivity alone does not guarantee trainability~\cite{mcclean2018,larocca2023,haug2021,larocca2022} and the quantum Fisher information matrix (QFIM)~\cite{braunstein1994} measures how the parameters populate the tangent space of the state~\cite{stokes2020}, and in quantum optimal control~\cite{koch2022,dalessandro2021}, where reachability is decided by Lie-algebraic criteria~\cite{jurdjevic1972,albertini2003} and cost by minimum-time bounds~\cite{khaneja2001,deffner2017}. The collision model lets both viewpoints be worked out explicitly, with the physical couplings as a two-dimensional family of drifts and the phases as controls.

The support structure of the collision map decides what can be reached. At the sampled generic collision angles the projective orbit of the localized input fills the \(2(C-1)\)-dimensional state space of the sector, where \(C=\binom{N}{K}\), in every sector for which the certificate was computed, that is, for \(C\leq21\); in sectors with \(C\) up to 1716 the optimizations reach fidelities above \(0.999\) for Dicke and Haar-random targets, which is consistent with full reachability but does not certify it. At special angles the collisions become permutations or the identity, conserved register occupations confine the dynamics to coordinate sectors, and the fidelity with any target is bounded by the weight of that target on the invariant support; for the Dicke state the bound is explicit and is attained numerically. Such an obstruction also leaves a quantitative footprint in the controllable region around it, because the number of rounds needed to exceed the ceiling grows at least as the inverse distance from the singular angle.

The cost of preparation then depends on where the couplings lie. At the four sampled generic couplings the number of rounds at which a given Dicke fidelity is first reached, expressed as the ratio of independent phase controls to the projective dimension, is nearly the same across 24 system--coupling pairs with \(C\) between 70 and 1716, and the QFIM shows that the same maximal tangent rank can carry different spectral conditioning; at the generic coupling the spectrum is close to that of a random Gaussian Jacobian. Near full SWAP the sequence reduces to a permutation backbone with weak exchanges; the detuning radius defines a rescaled time, the detuning direction selects an effective control problem, and the independently optimized solution of that problem predicts the microscopic number of rounds to within 10\% in all 36 tested cases. This slow-time law is the sharp form, at the full-SWAP corner, of the inverse-distance bound. A sweep of the detuning along two rays connects the two regimes. The Dicke horizon follows the slow-time law up to \(\varepsilon=0.1\), where the Dicke state reaches \(F=0.98\) in 7 or 8 rounds in sectors with \(C=70\), 126, and 252, against 14, 22, and 38 rounds at a generic coupling, and beyond \(\varepsilon\approx0.3\) every target returns to the generic cost. The slow-time value itself, computed from the effective model, is nearly the same for \(C\) from 35 to 252. In the same structured region the target matters: phase-uniform, random-sign, and Haar-random targets are delayed for every sampled instance. The effective exchange graph, which annihilates the Dicke state, offers an interpretation of this ordering.

We use Dicke states as the principal target, because their structure is what makes the near-SWAP regime favorable and because they connect this work to Refs.~\cite{vu2026pra,vu2026data,vu2026sd}; the reachability results, the control-budget organization, and the effective near-SWAP theory are formulated for an arbitrary target in the sector, the detuning sweep connects the near-SWAP and generic regimes along two rays, and the target-dependence study compares Dicke states with four ensembles of generic targets. The results have different standing, which we label throughout: the invariant-sector ceilings, the inverse-distance bound, and the slow-time theorem are proved; the projective-orbit certificates are numerical; the control-budget organization, the crossover, and the target ordering are empirical; the exchange-graph mechanism is an interpretation. All results refer to the fixed-excitation sector reachable from the \(K\)-excitation input, to the shuttle--register architecture defined in Sec.~\ref{sec:model}, and to the system sizes stated with each result. Sections~\ref{sec:reach} and \ref{sec:qfim} treat reachability and the QFIM geometry, Sec.~\ref{sec:results} the preparation cost in the two regimes and the target dependence, and Sec.~\ref{sec:discussion} the implications and limitations.

\section{Collision model with interleaved phase control}
\label{sec:model}

\subsection{Collision rounds}
\label{subsec:passive-map}

The register consists of \(N\) qubits, of which \(K\) act as shuttles and \(n=N-K\) form two chains. All interactions are partial SWAPs,
\begin{equation}
    U(\gamma)=\cos\gamma\,I+i\sin\gamma\,\SWAP=e^{i\gamma\,\SWAP},
    \label{eq:partial-swap-intro}
\end{equation}
where the second equality follows from \(\SWAP^2=I\). Since \(\SWAP=\tfrac12(I+XX+YY+ZZ)\) and the three Pauli products commute,
\begin{equation}
    e^{i\gamma\,\SWAP}=e^{i\gamma/2}\,e^{i\gamma(XX+YY)/2}\,e^{i\gamma ZZ/2},
    \label{eq:heisenberg-split}
\end{equation}
so the collision is an isotropic Heisenberg exchange: an excitation-exchange factor of iSWAP type together with a conditional phase. Within the \(K\)-excitation sector the \(ZZ\) factor is a two-body phase, which depends on whether the two sites are equally occupied, and the local \(\RZ\) layers introduced below cannot remove it. On hardware the gate corresponds to an exchange interaction of iSWAP or fSim type combined with a conditional phase; in a CNOT/CZ gate basis each partial SWAP compiles into three entangling gates~\cite{vu2026data}. On platforms whose native interaction is a pure \(XY\) exchange the conditional phase is absent, and the reachability and cost statements below would have to be rederived for that model. The singular angles \(\gamma\in\{0,\pi/2,\pi\}\), at which the collision is a permutation up to a phase, give \(I\), \(i\cdot\SWAP\), and \(-I\), with \(i\cdot\SWAP\) denoting scalar multiplication of the SWAP operator by \(i\). The interaction conserves the number of excitations, so we work in the \(K\)-excitation subspace, whose dimension is
\begin{equation}
    C=\binom{N}{K}.
\end{equation}
We refer to it as the \(\sect{N}{K}\) sector and write \(\Dstate{N}{K}\) for its Dicke state.

\begin{figure*}[t]
\centering
\textbf{(a)}\\[-2pt]
\includegraphics[width=0.58\textwidth]{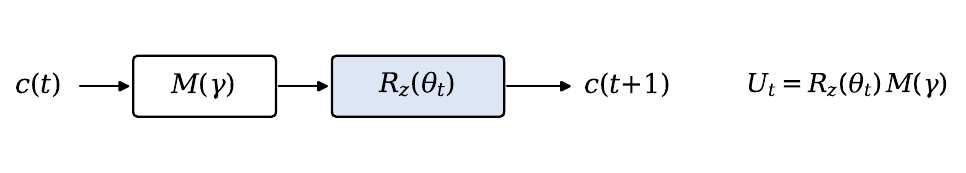}\\[5pt]
\textbf{(b)}\\[-2pt]
\includegraphics[width=0.90\textwidth]{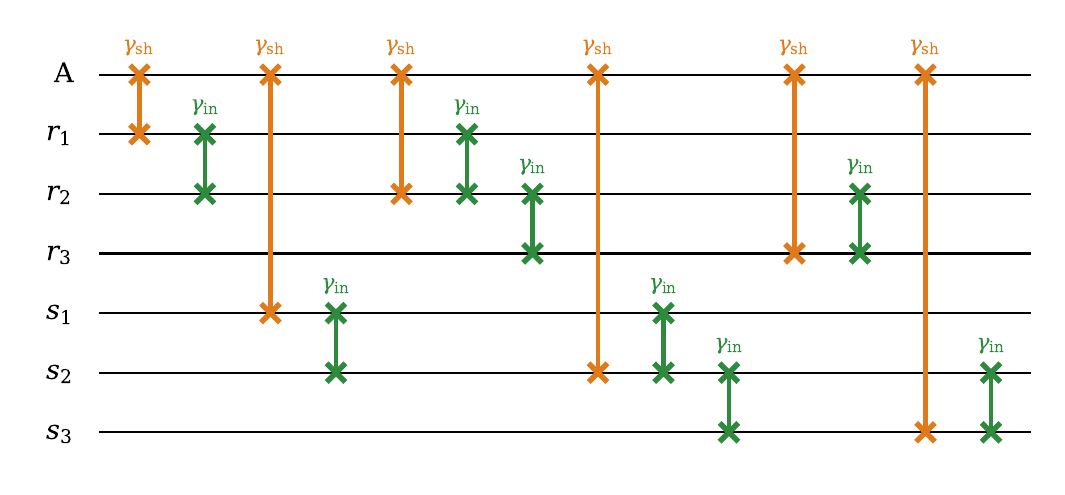}
\caption{Collision model with interleaved phase control in the \(K\)-excitation sector. (a)~One controlled round, \(U_t=\RZ(\theta_t)M(\gam)\): a full collision round at fixed angles \((\gamin,\gamsh)\) is followed by a layer of local \(\RZ\) rotations with free angles. (b)~One shuttle subround within \(M(\gam)\), following Refs.~\cite{cakmak2019,vu2026pra}. A shuttle qubit \(A\) alternates between the chains \(r\) and \(s\). At each visited site it performs a shuttle--register partial SWAP with angle \(\gamsh\) (orange), followed by intra-register partial SWAPs with angle \(\gamin\) (green) between that site and its nearest neighbors in the same chain. Each partial SWAP is drawn as two \(\times\) markers joined by a vertical line. The subround is repeated for all \(K\) shuttles. The collision angles stay fixed during a run, and the optimizer varies the \(T\times N\) phase angles, of which \(T(N-1)\) are independent. The site labels in the figure are one-based (\(r_1,s_1,\ldots\)); Appendix~\ref{app:schedule} uses zero-based indices.}
\label{fig:schematic}
\end{figure*}

Let \(c(t)\in\mathbb{C}^C\) denote the amplitude vector in the computational basis of the sector. Following Refs.~\cite{cakmak2019,vu2026pra}, each shuttle qubit is swept through the two chains: a shuttle--register partial SWAP with angle \(\gamsh\) is followed by intra-register partial SWAPs with angle \(\gamin\) between the visited qubit and its nearest neighbors, and the shuttle alternates between the two chains as the sweep proceeds. One sweep of one shuttle is a subround, and a round consists of the \(K\) subrounds of all shuttles, as in Refs.~\cite{vu2026pra,vu2026sd}. One round acts as the unitary stroboscopic map
\begin{equation}
    c(t+1)=M(\gam)c(t),\qquad \gam=(\gamin,\gamsh).
    \label{eq:passive-map}
\end{equation}
If the chronological schedule of the round is \(e_\ell=(a_\ell,b_\ell)\) with family \(\sigma_\ell\in\{\mathrm{in},\mathrm{sh}\}\), then
\begin{equation}
\begin{aligned}
 M(\gamin,\gamsh)
 &=\overleftarrow{\prod_{\ell=1}^{L}}
 e^{i\gamma_{\sigma_\ell}S_{a_\ell b_\ell}},\qquad
 S_{ab}=\SWAP_{ab}.
\end{aligned}
\label{eq:ordered-collision-word}
\end{equation}
Collision 1 acts first, so the rightmost factor is the earliest interaction. Appendix~\ref{app:schedule} specifies the site indices, the neighbor sets, and the complete order. Figure~\ref{fig:schematic} shows one round.

The shuttle qubits remain part of the \(N\)-qubit state throughout. The initial state is the computational-basis state with the \(K\) excitations on the shuttle sites. For the Dicke target, the fidelity is
\begin{equation}
    F(c)=\frac{1}{C}\left|\sum_{s=1}^{C}c_s\right|^2,
    \label{eq:fidelity}
\end{equation}
so \(F=1\) when all amplitudes have the same magnitude and phase; for a general target \(\ket{\psi_{\rm tar}}\) of the sector, \(F=|\langle\psi_{\rm tar}|c\rangle|^2\). In the passive protocol of Ref.~\cite{vu2026pra}, \((\gamin,\gamsh)\) are the only adjustable parameters, the dynamics can be quasiperiodic, and the attainable fidelity depends strongly on the chosen pair.

\subsection{Phase layers and gauge reduction}
\label{subsec:rz-control}

Let \(P\in\{0,1\}^{C\times N}\) be the occupation matrix of the sector basis, with \(P_{sj}=1\) when basis state \(s\) has an excitation on qubit \(j\). After round \(t\) we apply
\begin{equation}
    \RZ(\theta)=e^{-i\theta Z/2}=
    \begin{pmatrix}e^{-i\theta/2}&0\\0&e^{i\theta/2}\end{pmatrix}
\end{equation}
to each qubit, with angles \(\theta_t=(\theta_{t1},\ldots,\theta_{tN})^{\mathsf T}\in\mathbb{R}^N\). Restricted to the sector, the layer is the diagonal unitary
\begin{equation}
    \bigl[\RZ(\theta_t)\bigr]_{ss}=
    e^{\,i\left[(P\theta_t)_s-\frac12\mathbf{1}_N^{\mathsf T}\theta_t\right]},
    \qquad s=1,\ldots,C.
    \label{eq:rz-layer}
\end{equation}
The second term in the exponent is independent of \(s\) and keeps the tensor-product convention of the single-qubit gates.

Every basis state of the sector contains \(K\) excitations, hence \(P\mathbf{1}_N=K\mathbf{1}_C\), and a uniform shift of all angles in one layer gives
\begin{equation}
    \RZ(\theta_t+\alpha\mathbf{1}_N)
    =e^{i\alpha(K-N/2)}\RZ(\theta_t).
    \label{eq:rz-gauge}
\end{equation}
This factor is a global phase and therefore an exact null direction of the pure-state QFIM. For \(1\leq K\leq N-1\) it is the only redundancy within one layer: with
\begin{equation}
    \Pi_\perp=I_C-\frac{\mathbf{1}_C\mathbf{1}_C^{\mathsf T}}{C},
\end{equation}
one has
\begin{equation}
    \ker(\Pi_\perp P)=\operatorname{span}\{\mathbf{1}_N\},\qquad
    \rank(\Pi_\perp P)=N-1.
\label{eq:occupation-rank}
\end{equation}

The controlled evolution is
\begin{equation}
    c(t+1)=\RZ(\theta_t)M(\gam)c(t),
    \qquad t=0,\ldots,T-1,
    \label{eq:controlled-map}
\end{equation}
with the raw parameter array \(\Theta=(\theta_0,\ldots,\theta_{T-1})\in\mathbb{R}^{T\times N}\). Removing one gauge direction per layer leaves \(T(N-1)\) independent phase controls. The collision map mixes amplitudes within the sector, and the phase layers steer the interference among them.

\subsection{Objective, optimization, and observed horizons}
\label{subsec:optimization}

We minimize the final infidelity
\begin{equation}
    \Loss(\Theta)=1-F(c(T)).
    \label{eq:loss}
\end{equation}
The map of Eq.~\eqref{eq:controlled-map} is differentiable in \(\Theta\), and the gradient of \(\Loss\) is obtained by automatic differentiation through the full trajectory. For each sampled horizon \(T\), the phase schedule is optimized from several independent restarts with Adam, followed where stated by L-BFGS; the restart counts, epochs, and learning rates of each study are listed in Appendix~\ref{app:methods}. Let \(F_{\rm best}(T)\) denote the best final fidelity among the restarts and let \(\mathcal H\) be the set of integer horizons searched in a study. For a reporting level \(F<1\), the optimizer-observed minimum horizon is
\begin{equation}
    \hat T_{\min}(F)=\min\{T\in\mathcal H:F_{\rm best}(T)\geq F\}.
    \label{eq:observed-hitting}
\end{equation}
The hat marks a finite numerical search. The intrinsic optimum \(T_{\min}\), used in the analytic results below, is the smallest horizon at which some control sequence reaches the level, so \(\hat T_{\min}(F)\geq T_{\min}(F)\) and the observed horizon is an upper bound. We keep the complete curves \(F_{\rm best}(T)\) whenever they are available and report several levels \(F\), or equivalently infidelities \(\eta=1-F\), to resolve the high-fidelity regime.

\section{Reachability}
\label{sec:reach}

\subsection{Projective reachability}
\label{subsec:lie-test}

The restricted \(\RZ\) generator on qubit \(j\) is \(D_j=P_j-\tfrac12 I\), where \(P_j\) projects onto the basis states with an excitation on qubit \(j\). Its trace in the sector is \(C(K/N-1/2)\), so projective motion is described more naturally by the traceless representative \(\widetilde D_j=P_j-(K/N)I\); the two differ by a central term and induce the same projective dynamics. With the skew-Hermitian generators \(i\widetilde D_j\), define the drift-conjugated algebra
\begin{equation}
    \mathfrak h(\gam)=\Lie\bigl\{iM(\gam)^k\widetilde D_jM(\gam)^{-k}\;\big|\;
    j=1,\ldots,N,\ k\in\mathbb Z\bigr\}.
\label{eq:lie}
\end{equation}
For a state \(\ket\psi\), this algebra generates the projective tangent space
\begin{equation}
    d_{\rm orb}(\gam)
    =\dim_{\mathbb R}\bigl\{(I-\ket\psi\!\bra\psi)X\ket\psi\;\big|\;
    X\in\mathfrak h(\gam)\bigr\}.
\label{eq:dorb}
\end{equation}
When \(d_{\rm orb}=2(C-1)\), the orbit fills the projective dimension of the sector. Because the dynamical group is compact, a full orbit dimension means that the orbit is the whole projective space, so every state of the sector is reachable exactly within the closure of the controlled dynamics, with a number of rounds that is unbounded in general~\cite{jurdjevic1972,albertini2003,dalessandro2021}. The numerical construction truncates the conjugation index \(k\) and the closure of the algebra, so the computed \(d_{\rm orb}\) is a lower bound on the orbit dimension, and a full value is a sufficient certificate. Exact support invariants complement the certificate by identifying couplings whose controlled trajectories stay in a proper coordinate sector. Appendix~\ref{app:lie-proof} sketches a generator argument and states the step that it leaves open, and Appendix~\ref{app:singular-surveys} reports the certificates computed for the sectors \(\sect{4}{1}\), \(\sect{5}{2}\), \(\sect{6}{2}\), \(\sect{6}{3}\), and \(\sect{7}{2}\), that is, for \(C\leq21\): the sampled generic couplings attain \(d_{\rm orb}=2(C-1)\) in each of them. For larger sectors the statement rests on the optimizations of Sec.~\ref{sec:results}, which reach fidelities above \(0.999\) for Dicke and Haar-random targets with \(C\) up to 1716; a certificate for those sectors has not been computed.

\subsection{Singular couplings and invariant supports}
\label{subsec:commensurate}

The partial-SWAP convention gives
\begin{equation}
    U(0)=I,\qquad U(\pi/2)=i\cdot\SWAP,\qquad U(\pi)=-I .
    \label{eq:commensurate-gates}
\end{equation}
We call these the singular angles, or permutation points, and write
\begin{equation}
    \Gamma_{\rm s}=\{0,\pi/2,\pi\},\qquad \mathcal{I}_\gamma=[0,\pi]
    \label{eq:gamma-c}
\end{equation}
for the set of singular angles and the full angle interval.
The singular structure is asymmetric in \(\gamin\) and \(\gamsh\), and two exact obstructions follow directly from the collision word.

If \(\gamsh\in\{0,\pi\}\), every shuttle--register collision is proportional to the identity. The shuttle is the only transport channel from the initially excited shuttle sites into the chains~\cite{cakmak2019,vu2026pra}, so at these angles the excitations remain on the shuttle sites for every \(\gamin\), and the trajectory stays on a single computational-basis ray.

A second exact obstruction appears when both angles belong to \(\Gamma_{\rm s}\). Every collision gate is then monomial and the diagonal phase layers preserve this structure. A localized input remains a single basis ray under permutations and phases, and its Dicke fidelity is fixed at
\begin{equation}
    F=\frac{1}{C}.
    \label{eq:monomial-bound}
\end{equation}
More generally, confinement to a coordinate support \(\Omega\) bounds the fidelity with an arbitrary target by the weight of that target on \(\Omega\), and the Dicke fidelity by \(|\Omega|/C\) (Appendix~\ref{app:singular-surveys}); the monomial case has \(|\Omega|=1\). The exact obstruction set is
\begin{equation}
    \mathcal{S}_{\rm exact}
    =\left(\mathcal{I}_\gamma\times\{0,\pi\}\right)
    \cup\left(\Gamma_{\rm s}\times\Gamma_{\rm s}\right)
\label{eq:exact-obstruction}
\end{equation}
in the \((\gamin,\gamsh)\) plane.

The line \(\gamin=\pi/2\) has an exact invariant-sector description. Because the local phase torus is diagonal, every common invariant subspace of the controlled dynamics is a coordinate subspace. At full intra-register SWAP, the collision backbone and the shuttle defects preserve one logical register occupation when \(n=3\) and two when \(n\geq4\). Starting from the shuttle-localized input, the dynamics is confined to a proper coordinate sector, which gives
\begin{equation}
 F_{\max}\leq
 \begin{cases}
 \dfrac{N-K}{N}, & n=3,\\[4pt]
 \dfrac{(N-K)(N-K-1)}{N(N-1)}, & n\geq4,
 \end{cases}
 \label{eq:fmax-comp}
\end{equation}
for the Dicke target. The two cases are the ratios \(|\Omega|/C\) of the preserved coordinate sectors. The confinement holds at every round and for every phase schedule, so Eq.~\eqref{eq:fmax-comp} bounds the fidelity at every horizon. Table~\ref{tab:appB} reports representative saturation data: the optimized fidelities reach the analytic ceilings in \(\sect{5}{2}\), \(\sect{6}{3}\), and \(\sect{7}{3}\), and in the orbit survey of Appendix~\ref{app:singular-surveys} the computed orbit dimension fills the projective dimension of each invariant support.

The defect first appears at \(n=3\), when intra-register collisions enter the schedule; for \(n=1\) and \(n=2\) the intra-register family is absent and the dynamics is independent of \(\gamin\). The all-input support calculation in \(\sect{6}{2}\) resolves four invariant sectors with dimensions \(6\), \(4\), \(4\), and \(1\). Appendix~\ref{app:component-proof} derives these sectors and the finite-exception qualification, and the accompanying tables report the saturation checks.

\subsection{Cost of leaving a singular surface}
\label{subsec:singular-cost-bound}

An exact obstruction is a statement about one point of the coupling plane, but it also constrains the controllable region around it. Let \(\gamma_\alpha^\star\) be a singular value of one collision family, let \(F_0\) be the fidelity ceiling at that value, and let \(L_\alpha\) be the number of occurrences of that family in one round. For any \(F>F_0\), the pure-state Fubini--Study distance \(d_{\rm FS}(\psi,\phi)=\arccos|\langle\psi|\phi\rangle|\)~\cite{anandan1990} between the singular and the perturbed trajectories gives
\begin{equation}
 T_{\min}(\gamma_\alpha;F)\geq
 \frac{\arccos\sqrt{F_0}-\arccos\sqrt F}
 {L_\alpha|\gamma_\alpha-\gamma_\alpha^\star|}.
 \label{eq:singular-boundary-speedlimit}
\end{equation}
The bound holds for every phase schedule, and for every target state with ceiling \(F_0\) on the singular surface. Moving a collision angle a small distance away from the surface restores the missing motion only gradually, because each collision can move the state by a Fubini--Study angle of at most \(|\gamma_\alpha-\gamma_\alpha^\star|\), and the required distance has to be accumulated over rounds. In the convention \(d_{\rm FS}=\arccos|\langle\psi|\phi\rangle|\) used here, the maximal speed of one collision per unit angle is one; Ref.~\cite{anandan1990} normalizes the same distance with an additional factor of 2. The right-hand side of Eq.~\eqref{eq:singular-boundary-speedlimit} vanishes as \(F\to F_0^{+}\), as it must, and it is positive for every \(F>F_0\). On the frozen-shuttle lines \(\gamsh\in\{0,\pi\}\), \(F_0=1/C\) and \(L_{\rm sh}=K(N-K)\), so a Dicke fidelity \(F\) requires \(T_{\min}\geq[\arccos C^{-1/2}-\arccos\sqrt F]/[K(N-K)\,|\gamsh-\gamma_{\rm sh}^\star|]\). On the intra-register full-SWAP line, \(L_{\rm in}=2K(N-K-2)\), so every Dicke fidelity above the ceiling of Eq.~\eqref{eq:fmax-comp} requires a number of rounds that grows at least as \(|\gamin-\pi/2|^{-1}\). Appendix~\ref{app:singular-speed} gives the distance argument and the collision-count audit. This line crosses the middle of the search domain of the passive protocol of Ref.~\cite{vu2026sd}, whose fidelity landscape is therefore capped exactly on it. Section~\ref{subsec:predictive} applies the same bound at the full-SWAP corner, where both families are detuned at once, and identifies the slow-time theory as its sharp form there.

\section{Quantum Fisher information geometry}
\label{sec:qfim}

\subsection{Rank and parameter redundancy}
\label{subsec:rank-why}

The reachability analysis determines which projective directions exist, but it does not show how a finite horizon populates them. After the per-layer gauge direction of Eq.~\eqref{eq:rz-gauge} is removed, a horizon of \(T\) rounds supplies \(T(N-1)\) independent phase controls, while the normalized output state lies in \(\cproj\) with real dimension
\begin{equation}
    \dim_{\mathbb{R}}\cproj=2(C-1).
    \label{eq:projective-dim}
\end{equation}
The number of independent tangent directions generated at horizon \(T\) is the rank of the QFIM of the output state with respect to the phase parameters~\cite{braunstein1994}. For the final state \(c(T)\), let \(\partial_\mu c=\partial c(T)/\partial\Theta_\mu\); then
\begin{equation}
    \qfim_{\mu\nu}=4\,\mathrm{Re}\!\Bigl[
    \langle\partial_\mu c|\partial_\nu c\rangle
    -\langle\partial_\mu c|c\rangle\langle c|\partial_\nu c\rangle
    \Bigr],
\label{eq:qfim}
\end{equation}
and the two counts above give the bound
\begin{equation}
 r_{\rm hard}\leq\min\{T(N-1),2(C-1)\},
 \label{eq:rank-bound}
\end{equation}
where the hard rank \(r_{\rm hard}\) counts eigenvalues above the relative cutoff specified in Appendix~\ref{app:methods}. In the overparametrization analysis of Larocca \textit{et al.}~\cite{larocca2023}, rank saturation marks the parameter count at which the accessible local dimension reaches its ceiling; in the fixed-excitation sector that ceiling is \(2(C-1)\). The parameter count \(T\times N\) and the number of independent directions therefore differ by the \(T\) gauge directions and, beyond saturation, by the excess of controls over the projective dimension.

\subsection{Spectral conditioning}
\label{subsec:conditioning}

The QFIM spectrum resolves how strongly the state responds along the populated directions. Besides \(r_{\rm hard}\) we use the participation ratio
\begin{equation}
    \reff=\frac{\left(\sum_k\lambda_k\right)^2}{\sum_k\lambda_k^2}.
    \label{eq:reff}
\end{equation}
When the same retained positive eigenvalues enter \(\reff\), \(r_{\rm hard}\), and their squared coefficient of variation \(\mathrm{CV}^2\),
\begin{equation}
    \reff=\frac{r_{\rm hard}}{1+\mathrm{CV}^2},
    \label{eq:pr-cv}
\end{equation}
and \(1\leq\reff\leq r_{\rm hard}\leq2(C-1)\). The participation ratio falls when the same tangent dimension is carried by a more uneven spectrum. The condition number
\begin{equation}
    \chi_F=\frac{\lambda_{\max}}{\lambda_{\min}^{+}},
    \label{eq:chiF}
\end{equation}
where \(\lambda_{\min}^{+}\) is the smallest eigenvalue above the hard-rank cutoff, resolves the weakest retained direction.

\subsection{Participation landscape}
\label{subsec:landscape}

\begin{figure}[t]
\centering
\includegraphics[width=0.95\columnwidth]{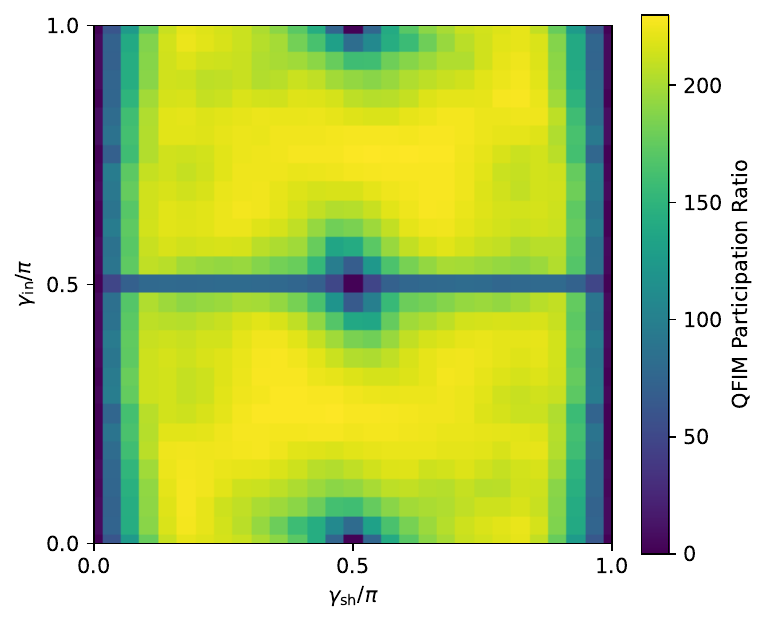}
\caption{Participation ratio of the exact QFIM for the half-filled sector \(\sect{10}{5}\) (\(C=252\)) over the coupling plane. Color encodes \(\reff\) on a \(29\times29\) grid over \(\gamin\) and \(\gamsh\) that includes \(\{0,\pi/2,\pi\}\) in each coordinate, evaluated at \(T=\lceil2(C-1)/(N-1)\rceil=56\). Each value uses the first of four stored phase vectors with entries drawn uniformly from \([-1,1]\); the remaining three enter a robustness check of the landscape across the grid (Appendix~\ref{app:methods}). The horizontal trench at \(\gamin=\pi/2\) is the invariant-sector obstruction of Eq.~\eqref{eq:fmax-comp}, and the vertical trenches at \(\gamsh\in\{0,\pi\}\) are frozen transport.}
\label{fig:effective-rank-landscape}
\end{figure}

Figure~\ref{fig:effective-rank-landscape} maps the participation ratio over the coupling plane for the half-filled sector \(\sect{10}{5}\) at the horizon at which the control count first reaches the projective dimension. The frozen-shuttle lines \(\gamsh=0\) and \(\gamsh=\pi\) appear as narrow low-\(\reff\) trenches, while the broader feature around \(\gamin=\pi/2\) follows the invariant-sector obstruction of Sec.~\ref{sec:reach}. The full-SWAP shuttle line \(\gamsh=\pi/2\) behaves differently: when the partner coupling is a nonsingular angle, its participation ratio returns to the generic range. The landscape thus distinguishes frozen transport, the intra-register defect, and the generic bulk through their local spectral geometry.

\subsection{Control ratio}
\label{subsec:kappa}

\begin{figure*}[t]
\centering
\begin{minipage}[t]{0.32\textwidth}
\centering
\textbf{(a)}\\[-1pt]
\includegraphics[width=\linewidth]{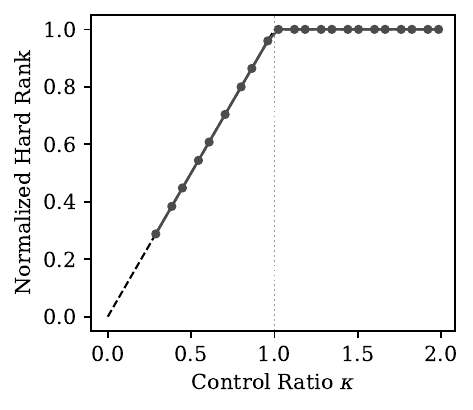}
\end{minipage}\hfill
\begin{minipage}[t]{0.32\textwidth}
\centering
\textbf{(b)}\\[-1pt]
\includegraphics[width=\linewidth]{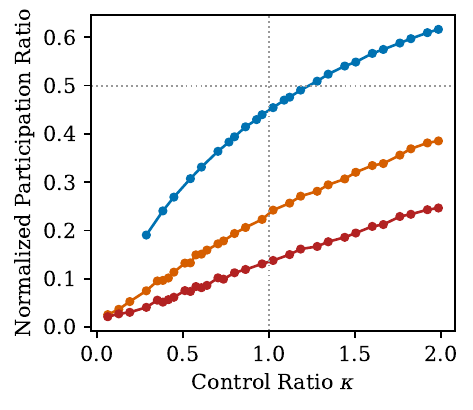}
\end{minipage}\hfill
\begin{minipage}[t]{0.32\textwidth}
\centering
\textbf{(c)}\\[-1pt]
\includegraphics[width=\linewidth]{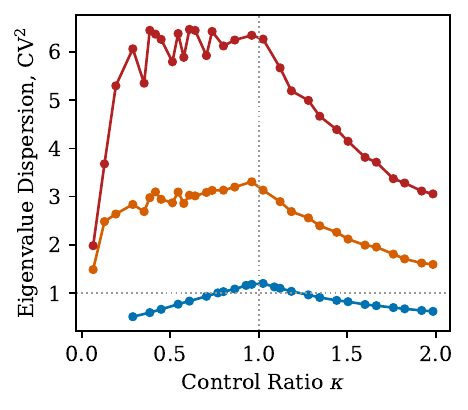}
\end{minipage}
\caption{Finite-horizon QFIM rank and conditioning for \(\sect{9}{4}\) (\(C=126\)), evaluated at the four Gaussian phase vectors (\(\sigma=0.25\)) of Appendix~\ref{app:methods} per horizon, as functions of the control ratio \(\kappa\). The generic coupling is sampled on horizons from \(T=9\) (\(\kappa\approx0.29\)) and the two near-SWAP couplings from \(T=2\) (\(\kappa\approx0.06\)), which is why the curves start at different \(\kappa\). (a)~Hard rank normalized by \(2(C-1)\): the three couplings \((\gamin,\gamsh)=(1,1)\), \((1.45,1.45)\), and \((1.5,1.5)\) give the same values, shown as one dark-gray curve, and follow the envelope \(\min(\kappa,1)\) (black dashed). (b)~Participation ratio normalized by \(2(C-1)\), and (c)~squared coefficient of variation of the retained spectrum; blue, orange, and red denote \((1,1)\), \((1.45,1.45)\), and \((1.5,1.5)\). The gray dotted vertical guide marks \(\kappa=1\), the gray dotted horizontal guide in (b) marks \(1/2\), and the black dashed curve in (c) marks \(\mathrm{CV}^2=1\); these are the Marchenko--Pastur values \(\kappa/(1+\kappa)\) and \(\min(\kappa,1/\kappa)\) at \(\kappa=1\). The same rank carries a more even spectrum at the generic coupling than near full SWAP.}
\label{fig:synthesis}
\end{figure*}

Figure~\ref{fig:synthesis}(a) shows finite-horizon rank growth. For \(\sect{9}{4}\), the exact QFIMs at the three displayed couplings follow
\begin{equation}
    r_{\rm hard}=\min\bigl[T(N-1),2(C-1)\bigr]
    \label{eq:rank-envelope}
\end{equation}
on the sampled horizon grid, so these couplings attain the bound of Eq.~\eqref{eq:rank-bound} at every sampled horizon. The first term is the number of independent phase controls and the second the projective ceiling. The invariant sectors of Sec.~\ref{sec:reach} have smaller supports and therefore smaller ceilings, as recorded in Appendix~\ref{app:singular-surveys}. The envelope suggests the dimensionless ratio
\begin{equation}
    \kappa(T)=\frac{T(N-1)}{2(C-1)}
    \label{eq:kappa}
\end{equation}
of independent controls to projective dimension; at \(\kappa=1\) the two counts are equal. Section~\ref{sec:results} places the observed horizons on this coordinate.

\subsection{Coupling dependence}
\label{subsec:synthesis}

The remaining panels of Fig.~\ref{fig:synthesis} show that the same hard rank can carry different spectral conditioning. In Fig.~\ref{fig:synthesis}(b), the generic spectrum distributes its weight more evenly than the near-SWAP spectra and reaches larger participation ratios at the same rank; Fig.~\ref{fig:synthesis}(c) shows the accompanying increase in eigenvalue dispersion. The sampled couplings share the maximal rank envelope while their local conditioning differs. At the generic coupling the spectrum is close to that of a random Gaussian Jacobian of shape \(2(C-1)\times T(N-1)\), whose Marchenko--Pastur statistics give \(\reff/[2(C-1)]\simeq\kappa/(1+\kappa)\) and \(\mathrm{CV}^2\simeq\min(\kappa,1/\kappa)\); the guides in Fig.~\ref{fig:synthesis}, \(\reff/[2(C-1)]=1/2\) and \(\mathrm{CV}^2=1\) at \(\kappa=1\), are these values at \(\kappa=1\), and the generic-coupling curves lie within about 10--15\% of these values over the sampled range. The physical reading is that the round map at a generic coupling scrambles the phase perturbations, so the tangent directions populate the sector like those of a random matrix, whereas near full SWAP the dynamics is structured and the spectrum is strongly anisotropic. The rank scale is common, which motivates \(\kappa\) as the coordinate of the cost analysis; the conditioning is coupling dependent, and Sec.~\ref{subsec:target-selectivity} shows that a less even spectrum can still be favorable for a target aligned with its dominant directions. Across the 72 cases of the pre-drawn survey of Sec.~\ref{subsec:kappa-collapse}, evaluated with the four Gaussian phase vectors of this figure at the horizons \(\kappa=0.5\) and 1, conditioning and cost are correlated with this sign: the rank correlation between \(\reff/[2(C-1)]\) and the observed \(\kappa^{(0.98)}\) is between \(0.53\) and \(0.60\) in both sectors. It comes mainly from the eleven cases with \(\kappa^{(0.98)}<0.6\), all with \(\gamsh\) within \(0.6\) of 0 or \(\pi\), whose spectra are the least even, with a mean \(\reff/[2(C-1)]\) of \(0.17\)--\(0.19\) at \(\kappa=0.5\) against \(0.26\) in the other cases; among the other cases, where \(\kappa^{(0.98)}\) takes only two or three values on the integer grid, the rank correlation is \(0.15\) to \(0.32\).

\section{Preparation cost}
\label{sec:results}

\subsection{Generic couplings}
\label{subsec:kappa-collapse}

The exact support ceilings of Sec.~\ref{sec:reach} define the first cost regime: at a singular coupling a target above the ceiling is unreachable, and Eq.~\eqref{eq:singular-boundary-speedlimit} makes \(T_{\min}\) grow at least as \(|\gamma_\alpha-\gamma_\alpha^\star|^{-1}\) nearby. Away from these surfaces, the sampled couplings recover the full projective-orbit certificate and follow the QFIM rank envelope. We therefore place the observed horizon \(\hat T_{\min}(F)\) on the control ratio \(\kappa\) of Eq.~\eqref{eq:kappa}, so that the rank scale and the reaching curves share one coordinate.

\begin{figure}[tbp]
\centering
\includegraphics[width=\columnwidth]{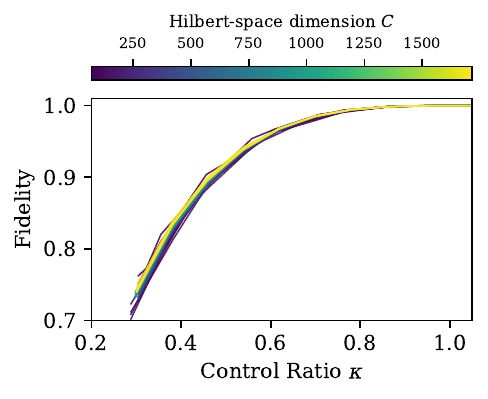}
\caption{Dicke fidelity at generic couplings as a function of the control ratio \(\kappa\) for 24 system--coupling curves with \(C=70\)--\(1716\). Each thin solid curve is a monotone best-over-restart envelope. The color bar encodes the sector dimension \(C\), from dark purple at \(C=70\) to yellow at \(C=1716\), so curves with equal or nearby \(C\) share a visual family. The curves of all sectors and couplings follow a common progression on \(\kappa\), and the budget needed for a given fidelity rises smoothly with the requested level.}
\label{fig:kappa-collapse}
\end{figure}

Figure~\ref{fig:kappa-collapse} collects 24 curves for the Dicke target from six sectors, \(\sect{8}{4}\) through \(\sect{13}{6}\), with the same four coupling pairs in each sector and \(C\) from 70 to 1716. Plotted against \(\kappa\), the best-fidelity envelopes of all sectors lie close together across the full fidelity range. At any sampled level \(F\), the observed crossing can be written as
\begin{equation}
    \hat T_{\min}(F)=
    \frac{2\kappa^{(F)}(C-1)}{N-1},
    \label{eq:generic-budget}
\end{equation}
where \(\kappa^{(F)}\) is the crossing budget at level \(F\). Equation~\eqref{eq:generic-budget} by itself is a change of variable; the empirical content of the figure is that \(\kappa^{(F)}\) is nearly independent of \((N,K)\) at these couplings. Across the 24 curves the median of \(\kappa^{(F)}\) is \(0.619\) at \(F=0.95\), with the individual values spread over \(0.558\)--\(0.627\); \(0.702\) (\(0.697\)--\(0.710\)) at \(F=0.98\); \(0.780\) (\(0.761\)--\(0.789\)) at \(F=0.99\); and \(0.941\) (\(0.864\)--\(0.964\)) at \(F=0.999\). Because \(T\) is an integer, one step in the horizon corresponds to a resolution \((N-1)/[2(C-1)]\) on this scale, between \(0.004\) and \(0.051\) in these sectors, which accounts for part of the spread. Equation~\eqref{eq:generic-budget} is an empirical statement about the tested systems and the four sampled couplings. The normalization by \(2(C-1)/(N-1)\) is suggested by the control count. Among alternatives, \(C/N\) and \(C/(N-1)\) collapse the 24 crossings equally well, with a coefficient of variation between \(0.4\%\) and \(2.6\%\) at the four levels for the three forms, close to the integer-horizon resolution; \(C/K\) and \(C/(N-K)\) give \(4\)--\(7\%\), and \(C-1\) and \(C^{1/2}\) give \(18\)--\(39\%\). The data therefore fix a cost proportional to \(C/N\) up to the resolution and do not distinguish between the forms with this scaling.

Finite-fidelity reaching can occur before the tangent space saturates. Reaching a fidelity neighborhood requires the endpoint map to intersect that neighborhood, whereas local surjectivity requires independent motion in every projective direction. The crossings in Fig.~\ref{fig:kappa-collapse} lie below the full-rank scale \(\kappa=1\) over most of the reported range, which makes the distinction visible on the common coordinate.

An independent survey uses 36 coupling pairs drawn and stored before any optimization, each coordinate at least \(0.15\) from the singular angles, in the sectors \(\sect{7}{3}\) and \(\sect{8}{4}\). All 72 coupling--sector cases reach the level \(F=0.98\) (Appendix~\ref{app:generic-sampling}). The bulk of the survey lies at \(\kappa=0.66\)--\(0.71\), in agreement with the campaign value, and the largest observed budgets, \(0.706\) and \(0.710\), coincide with it. The survey also shows coupling dependence: six of the 72 cases, from four of the 36 pairs, reach \(F=0.98\) at \(\kappa\) between \(0.25\) and \(0.46\), two to three times faster than the bulk. All four pairs have \(\gamsh\) within \(0.37\) of \(0\) or \(\pi\), where the shuttle collision is weak, and the fastest, \((\gamin,\gamsh)=(2.82,2.83)\), has both angles within \(0.32\) of \(\pi\). These weak-shuttle-collision couplings form a third regime that this paper does not analyze. The campaign value \(\kappa^{(F)}\) is therefore the plateau of the generic bulk, an upper envelope over couplings; Fig.~\ref{fig:kappa-collapse} resolves its fidelity dependence.

\subsection{Near full SWAP}
\label{subsec:predictive}

The behavior changes near the full-SWAP point \((\gamin,\gamsh)=(\pi/2,\pi/2)\), where each collision is almost a permutation. We use Cartesian components for the two detunings and polar coordinates for their distance and direction,
\begin{equation}
\begin{aligned}
    \bm\delta
    &=\begin{pmatrix}\delta_{\rm in}\\ \delta_{\rm sh}\end{pmatrix}
      =\begin{pmatrix}\pi/2-\gamin\\ \pi/2-\gamsh\end{pmatrix}
      =\varepsilon\hat{\bm v}(\phi),\\
    \hat{\bm v}(\phi)&=\begin{pmatrix}\cos\phi\\ \sin\phi\end{pmatrix},
    \qquad
    \varepsilon=\|\bm\delta\|_2 .
\end{aligned}
\label{eq:radial-detuning}
\end{equation}
Here \(\varepsilon\) is the Euclidean distance from full SWAP and \(\phi\) labels the ray, so \(\gamin=\pi/2-\varepsilon\cos\phi\) and \(\gamsh=\pi/2-\varepsilon\sin\phi\). Each collision factorizes exactly as
\begin{equation}
    U\!\left(\frac\pi2-\varepsilon v\right)
    =i\cdot\SWAP\,e^{-i\varepsilon v\,\SWAP}.
    \label{eq:nearcomm-collision}
\end{equation}
The order of the collisions determines the weak generator. Write \(S_\ell=S_{a_\ell b_\ell}\) and define the permutation prefixes and pulled-back swaps
\begin{equation}
\begin{aligned}
 Q_0&=I,\qquad Q_\ell=S_\ell\cdots S_1,\\
 \widetilde S_\ell&=Q_{\ell-1}^{-1}S_\ell Q_{\ell-1},\qquad
 \Pi=Q_L .
\end{aligned}
\label{eq:defect-prefixes}
\end{equation}
Moving the permutation factors to the left gives the exact factorization
\begin{equation}
\begin{aligned}
 M(\varepsilon)&=i^L\Pi E_\varepsilon,\qquad
 E_\varepsilon=\overleftarrow{\prod_{\ell=1}^{L}}
 e^{-i\varepsilon\hat v_{\sigma_\ell}\widetilde S_\ell},\\
 H_\alpha&=\sum_{\ell:\,\sigma_\ell=\alpha}\widetilde S_\ell,
 \qquad \alpha\in\{\mathrm{in},\mathrm{sh}\}.
\end{aligned}
\label{eq:exact-defect-factorization}
\end{equation}
Each term is a SWAP on relabeled sites, repeated occurrences add their weights, and the two collision families define a weighted exchange graph. The factorization is exact along any ray; the rescaled-time reduction below becomes predictive when \(\varepsilon\) is small enough that the accumulated \(T\varepsilon^2\) remainder vanishes at fixed \(\tau=\varepsilon T\). In the generic interior this weak-defect ordering is absent, and the control ratio \(\kappa\) remains the relevant coordinate. The ordered round has the expansion
\begin{equation}
\begin{aligned}
    M\!\left(\tfrac\pi2\mathbf{1}-\varepsilon\hat{\bm v}\right)
    &=e^{i\chi}\Pi
    \left[I-i\varepsilon H_{\hat{\bm v}}+O(\varepsilon^2)\right],\\
    H_{\hat{\bm v}}
    &=\hat v_{\rm in}H_{\rm in}+\hat v_{\rm sh}H_{\rm sh}.
\end{aligned}
\label{eq:nearcomm-round}
\end{equation}

The inverse-detuning regime requires a multi-round description with the phase controls kept explicit. Let \(D_t=\RZ(\theta_t)\) and let \(W_t\) be the product of the first \(t\) controlled permutation rounds, with \(W_0=I\). In this interaction frame the weak generator at round \(t\) is
\begin{equation}
 H_t=W_t^{-1}H_{\hat{\bm v}}W_t .
 \label{eq:controlled-weak-generator}
\end{equation}
After removing the common phase and the final factor \(W_T\), the exact controlled product differs from the ordered product of the \(e^{-i\varepsilon H_t}\) by at most
\begin{equation}
 \|\mathcal U_{\rm int}^{(T)}-\mathcal V^{(T)}\|
 \leq T\varepsilon^2\Lambda^2,
 \qquad \Lambda=\sum_{\ell=1}^{L}|\hat v_{\sigma_\ell}| .
 \label{eq:uniform-control-error}
\end{equation}
The bound is uniform over the phase sequence, including controls that vary with \(\varepsilon\). At fixed system size it vanishes when \(\tau=\varepsilon T\) stays bounded, which yields a controlled continuous evolution in the rescaled time \(\tau\). Equation~\eqref{eq:uniform-control-error} is a worst case. The prefactor \(\Lambda^2\) is of order \(L^2\sim K^2n^2\), so the condition for the remainder to be small at fixed \(\tau\) is \(\varepsilon\ll(\tau\Lambda^2)^{-1}\), which tightens with system size, and at the detunings used for the validation in Sec.~\ref{subsec:eff-vs-micro} the bound itself exceeds unity: for \(\sect{8}{4}\) at \(\delta_{\rm sh}=0.00625\) on the ray \(q=1\), \(T\varepsilon^2\Lambda^2\approx3\) at \(\tau=0.7\), whereas the accumulated error of the reduction evaluated directly is \(0.05\) and the agreement of the hitting times is at the percent level. The product-formula remainder can be tightened by a factor \(1/2\), or replaced by a commutator-based bound, without changing the scaling in \(\varepsilon\); the evidence that the reduction is accurate at the validation detunings is the direct comparison of Sec.~\ref{subsec:eff-vs-micro}.

The resulting effective control problem retains both the phase freedom and the cyclic permutation order; a slow drift steered by fast local controls is the setting of time-optimal control in spin systems~\cite{khaneja2001}. Let \(\tau_*(\hat{\bm v},F)\) be its intrinsic hitting time for fidelity \(F\). Whenever this time is finite and the level is regular in the sense of Appendix~\ref{app:slow-time}, the intrinsic minimum number of rounds satisfies
\begin{equation}
 T_{\min}(\varepsilon;\hat{\bm v},F)
 =\frac{\tau_*(\hat{\bm v},F)+o(1)}{\varepsilon}.
 \label{eq:intrinsic-slow-time}
\end{equation}
Appendix~\ref{app:slow-time} gives the effective control set \(\mathcal K_{\hat{\bm v}}\), the convergence argument, and the regularity condition. At fixed \((N,K)\), the radius \(\varepsilon\) maps rounds to rescaled time, while the direction \(\hat{\bm v}\) selects \(\mathcal K_{\hat{\bm v}}\). The coefficient \(\tau_*(\hat{\bm v},F)\) is the hitting time of this direction-resolved problem and carries the target dependence; Eq.~\eqref{eq:intrinsic-slow-time} is a statement about the scaling with \(\varepsilon\) at fixed system size and does not give \(\tau_*\) in closed form.

The slow-time law is the sharp form of the inverse-distance bound of Sec.~\ref{subsec:singular-cost-bound} at this corner. The full-SWAP point \((\pi/2,\pi/2)\) belongs to \(\Gamma_{\rm s}\times\Gamma_{\rm s}\), where every collision is a permutation and the Dicke fidelity of a localized input is \(F_0=1/C\). Detuning both families at once by \(\varepsilon\hat{\bm v}\) moves the endpoint, over \(T\) rounds, by a Fubini--Study angle of at most \(T\varepsilon\sum_\ell|\hat v_{\sigma_\ell}|=T\varepsilon\Lambda\), and the argument of Appendix~\ref{app:singular-speed} gives
\begin{equation}
 \tau_*(\hat{\bm v},F)\geq
 \frac{\arccos C^{-1/2}-\arccos\sqrt F}{\Lambda}
 \label{eq:corner-bound}
\end{equation}
for the Dicke target. Equation~\eqref{eq:singular-boundary-speedlimit} states that \(\varepsilon T_{\min}\) stays above this value, and Eq.~\eqref{eq:intrinsic-slow-time} states that \(\varepsilon T_{\min}\) converges to a finite \(\tau_*\); the slow-time theory supplies the constant that the inverse-distance bound leaves open, and the divergence of \(T_{\min}\) as \(\varepsilon\to0\) is the monomial obstruction seen from the controllable side. The bound is far from tight, since for \(\sect{8}{4}\) on the ray \(q=1\) it gives \(\tau_*\geq0.06\) at \(F=0.98\), well below the crossings of Fig.~\ref{fig:eff-micro}, but it is exact and it is the same inequality in both regimes.

The directional dependence is visible in the effective control body. For the rays \(\hat{\bm v}\propto(1,2)\) and \((2,1)\), exact pulled-back transposition counts give different normalized edge-width profiles in \(\sect{6}{3}\) and \(\sect{7}{3}\). More generally, the intra-register and shuttle width vectors are linearly independent throughout the main regime \(N\geq5\), \(2\leq K\leq\lfloor N/2\rfloor\), so distinct positive rays produce effective control bodies that are not homothetic. Appendix~\ref{app:directional-body} gives the support-width argument and the deterministic bookkeeping check over all 840 sectors of this regime with \(N\leq60\).

The same geometry produces a target-specific angular cost. A protected endpoint occupation can change only through the intra-register defect, which gives, for the Dicke target,
\begin{equation}
 \tau_*^D(v,F)\geq
 \frac{[\arcsin\sqrt{K/N}-\arccos\sqrt F]_+}
 {2K|v_{\rm in}|}.
 \label{eq:endpoint-cut-bound}
\end{equation}
For \(F>1-K/N\), the Dicke hitting time therefore diverges as the intra-register component \(v_{\rm in}\) vanishes. The radius sets the overall scale, while the direction sets a different control geometry and can itself approach a singular cost boundary.

A finite-radius survey provides an independent scaling check. Across 50 sector--ray cells from \(\sect{5}{2}\) through \(\sect{10}{5}\), the quantity \(\tau_{\rm obs}=\varepsilon\hat T_{\min}(0.98)\) changes little as the radius decreases: the largest per-ray coefficient of variation is \(0.026\), and every fitted log--log slope lies within \(0.047\) of \(-1\). The coefficient varies systematically with direction, in agreement with the directional effective body.

\subsection{Effective theory against the microscopic model}
\label{subsec:eff-vs-micro}

\begin{figure}[t]
\centering
\includegraphics[width=0.98\columnwidth]{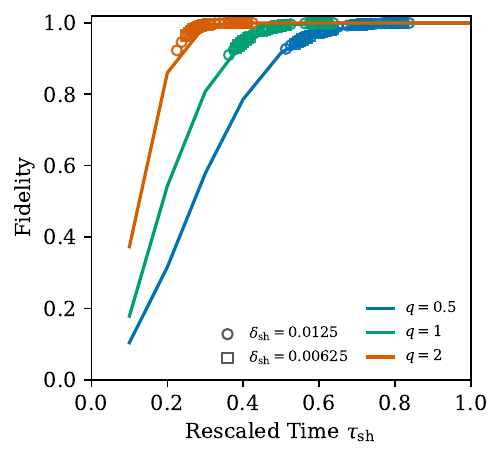}
\caption{Effective theory against the microscopic model near full SWAP for \(\sect{8}{4}\) (\(C=70\)). The horizontal axis is the component rescaled time \(\tau_{\rm sh}=\delta_{\rm sh}T\) and the vertical axis the Dicke fidelity. Solid blue, green, and orange curves are the independently optimized effective value functions for the ray ratios \(q=0.5\), \(1\), and \(2\). Open circles are microscopic partial-SWAP optimizations at \(\delta_{\rm sh}=0.0125\) and open squares at \(\delta_{\rm sh}=0.00625\); marker colors follow the \(q\) encoding of the curves. The microscopic points approach the effective curves as the detuning decreases. The \(\sect{6}{3}\) and \(\sect{7}{3}\) comparisons and the agreement table are given in Appendix~\ref{app:eff-micro}.}
\label{fig:eff-micro}
\end{figure}

The effective problem can be compared directly with the microscopic collision model. For this validation we use the ray ratio \(q=\delta_{\rm in}/\delta_{\rm sh}\in\{0.5,1,2\}\) and the component rescaled time \(\tau_{\rm sh}=\delta_{\rm sh}T\). We first solve the effective curves independently for \(\sect{6}{3}\), \(\sect{8}{4}\), and \(\sect{7}{3}\), then optimize the exact partial-SWAP dynamics at decreasing \(\delta_{\rm sh}\). Along a fixed ray, \(\varepsilon=\delta_{\rm sh}\sqrt{1+q^2}\) and \(\tau=\tau_{\rm sh}\sqrt{1+q^2}\), so the component and radial descriptions differ by one fixed factor. Figure~\ref{fig:eff-micro} shows the \(\sect{8}{4}\) comparison.

Across the three sectors, three directions, and four reported fidelities \(F\in\{0.95,0.98,0.99,0.999\}\), the smaller of the two microscopic detunings gives 36 comparisons. The normalized microscopic crossing \(\delta_{\rm sh}\hat T_{\min}^{\rm micro}(F)\) stays within \(10\%\) of the independently optimized effective crossing \(\hat\tau_{\min}^{\rm eff}(F)\) in all 36 cases, with most discrepancies at a few percent (Table~\ref{tab:eff-micro-agreement}). Multiplying both quantities by \(\sqrt{1+q^2}\) gives the radial comparison and leaves every relative discrepancy unchanged. The effective problem is solved without any input from the microscopic runs, so the agreement is a prediction of the near-SWAP theory and involves no fitted parameter.

\subsection{Crossover along a detuning ray}
\label{subsec:crossover}

\begin{figure}[t]
\centering
\includegraphics[width=0.98\columnwidth]{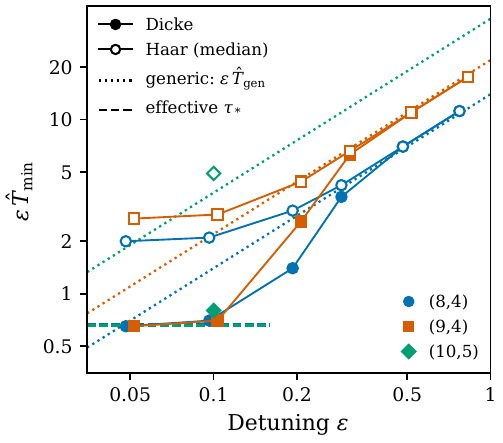}
\caption{Crossover from the near-SWAP regime to the generic regime along the ray \(q=1\), \(\gamin=\gamsh=\pi/2-\varepsilon/\sqrt2\), at \(F=0.98\). The vertical axis is \(\varepsilon\hat T_{\min}\), which is constant in the slow-time regime. Filled markers are the Dicke target and open markers the median of four Haar-random targets, with bars spanning their range; blue circles are \(\sect{8}{4}\) and orange squares \(\sect{9}{4}\), shifted slightly in \(\varepsilon\) for visibility. Green diamonds are the Dicke target and the median of 16 Haar targets for \(\sect{10}{5}\) at \(\varepsilon=0.10\), taken from the target study of Fig.~\ref{fig:target-specificity}. Dotted lines are \(\varepsilon\) times the generic horizons of 14, 22, and 38 rounds, and the dashed lines, which nearly coincide, are the slow-time values \(\tau_*=0.665\), \(0.658\), and \(0.655\) for \(\sect{8}{4}\), \(\sect{9}{4}\), and \(\sect{10}{5}\) from the effective problem, solved without microscopic input.}
\label{fig:crossover}
\end{figure}

The validation above uses detunings \(\varepsilon\leq0.028\), and the generic campaign uses couplings far from every singular angle. Figure~\ref{fig:crossover} connects the two regimes along the ray \(q=1\), \(\gamin=\gamsh=\pi/2-\varepsilon/\sqrt2\), for \(\varepsilon\) between \(0.05\) and \(0.8\) in \(\sect{8}{4}\) and \(\sect{9}{4}\). The coupling at \(\varepsilon=0.10\) is the coupling \((1.5,1.5)\) of Sec.~\ref{subsec:target-selectivity}, and the coupling at \(\varepsilon=0.8\) is \((1.005,1.005)\), next to the generic coupling \((1,1)\). The Dicke target and four Haar-random targets are optimized at every integer horizon up to a fixed cap (Appendix~\ref{app:sweep}). The figure shows \(\varepsilon\hat T_{\min}\) at \(F=0.98\), which is constant in the slow-time regime and grows linearly with \(\varepsilon\) at a fixed horizon. A second ray, \(q=2\), with \(\gamsh=\pi/2-\varepsilon/\sqrt5\) and \(\gamin=\pi/2-2\varepsilon/\sqrt5\), gives the same picture; Table~\ref{tab:sweep} lists both rays at all four fidelity levels.

Three regimes appear. For \(\varepsilon\geq0.5\) every target reaches \(F=0.98\) at the generic horizon of Fig.~\ref{fig:kappa-collapse}, 14 rounds in \(\sect{8}{4}\) and 22 in \(\sect{9}{4}\); at \(\varepsilon=0.3\) the Haar targets still do, and the Dicke target is one or two rounds faster. For \(\varepsilon\leq0.1\) the Dicke target follows the slow-time law: \(\varepsilon\hat T_{\min}\) is \(0.65\) at \(\varepsilon=0.05\) and \(0.70\) at \(\varepsilon=0.1\) in both sectors. The effective solutions, computed without microscopic input as in Sec.~\ref{subsec:eff-vs-micro}, give on this ray \(\tau_*=0.570\), \(0.665\), \(0.697\), and \(0.848\) at the four fidelity levels in \(\sect{8}{4}\) and \(0.565\), \(0.658\), \(0.689\), and \(0.826\) in \(\sect{9}{4}\); on the ray \(q=2\) they give \(0.597\), \(0.647\), \(0.664\), and \(0.867\), and \(0.585\), \(0.641\), \(0.659\), and \(0.843\). For \(F\leq0.99\) the microscopic Dicke horizon at \(\varepsilon=0.05\) and \(0.1\) lies within \(1.5\) rounds of \(\tau_*/\varepsilon\) in all 24 comparisons over the two sectors and the two rays. At \(F=0.999\) it agrees within two rounds at \(\varepsilon=0.05\), but at \(\varepsilon=0.1\) it exceeds the prediction by \(3.7\) and \(4.6\) rounds in \(\sect{9}{4}\), so the highest fidelity level is the first to show the finite detuning. The Haar targets are slow in the same range. At \(\varepsilon=0.1\) they need \(\varepsilon\hat T_{\min}=2.0\)--\(2.3\) in \(\sect{8}{4}\) and \(2.8\)--\(3.0\) in \(\sect{9}{4}\) on both rays, three to four times the Dicke value, and at \(\varepsilon=0.05\) they need 38 to 44 rounds in \(\sect{8}{4}\) and 52 to 54 in \(\sect{9}{4}\). Over both detunings and both rays \(\varepsilon\hat T_{\min}\) for the Haar targets stays between \(1.9\) and \(2.3\) in \(\sect{8}{4}\) and between \(2.6\) and \(3.0\) in \(\sect{9}{4}\), so the Haar targets also follow a slow-time law, with a slow-time value three to four times that of the Dicke target, and this value grows with the sector. In between, the Haar horizon joins the generic one at \(\varepsilon\approx0.2\), whereas the Dicke horizon rises from its slow-time value to the generic one between \(\varepsilon=0.1\) and \(0.5\). Among the sampled detunings the Dicke target is fastest at \(\varepsilon=0.1\) in \(\sect{9}{4}\), with 7 rounds on both rays, one third of the generic horizon, and at \(\varepsilon=0.1\)--\(0.2\) in \(\sect{8}{4}\), with 6 or 7 rounds, about half of the generic horizon.

Two consequences matter for what follows. First, Eq.~\eqref{eq:uniform-control-error} gives no control at \(\varepsilon=0.10\), but the Dicke horizon there is still the slow-time value, so the coupling \((1.5,1.5)\) lies at the edge of the slow-time regime for the Dicke target, while the Haar targets there are already slower than at a generic coupling. The edge depends on the sector. In \(\sect{10}{5}\) the effective solution gives \(\tau_*=0.562\), \(0.655\), \(0.690\), and \(0.836\) on this ray, and the Dicke horizons of the target study at \(\varepsilon=0.10\), 7, 8, 9, and 18 rounds, lie \(1.4\) to \(2.1\) rounds above \(\tau_*/\varepsilon\) for \(F\leq0.99\) and about ten rounds above it at \(F=0.999\), as expected from a remainder that grows with the number of collisions per round. Second, in the slow-time regime the Dicke horizon hardly depends on the sector: at \(\varepsilon\leq0.1\) it differs by at most one round between \(\sect{8}{4}\) and \(\sect{9}{4}\) for \(F\leq0.99\) on both rays, and by up to four rounds at \(F=0.999\), and the target study below gives 8 rounds at \(\varepsilon=0.10\) in \(\sect{10}{5}\), whereas the generic horizon grows from 14 to 22 and 38 rounds. The effective model shows the same in the limit \(\varepsilon\to0\): at \(F=0.98\) the slow-time value \(\tau_*\) lies between \(0.64\) and \(0.73\) for \(\sect{7}{3}\), \(\sect{8}{4}\), and \(\sect{9}{4}\) on each of the rays \(q=0.5\), 1, and 2, and only the smallest sector, \(\sect{6}{3}\), has larger values, from \(0.80\) to \(1.05\). On the ray \(q=1\) the value for \(\sect{10}{5}\), \(0.655\), lies in the same range, so that \(\tau_*\) is nearly constant from \(C=35\) to \(C=252\) on that ray. The Haar horizon at \(\varepsilon=0.10\) grows with the sector like the generic one.

\subsection{Target dependence}
\label{subsec:target-selectivity}

\begin{figure*}[t!]
\centering
\begin{minipage}[t]{0.485\textwidth}
\centering
\textbf{(a)}\\[-1pt]
\includegraphics[width=\linewidth]{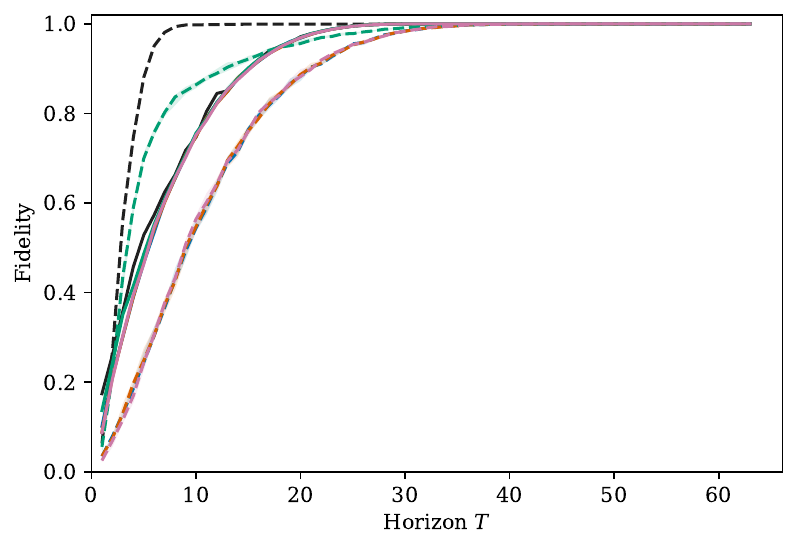}
\end{minipage}\hfill
\begin{minipage}[t]{0.485\textwidth}
\centering
\textbf{(b)}\\[-1pt]
\includegraphics[width=\linewidth]{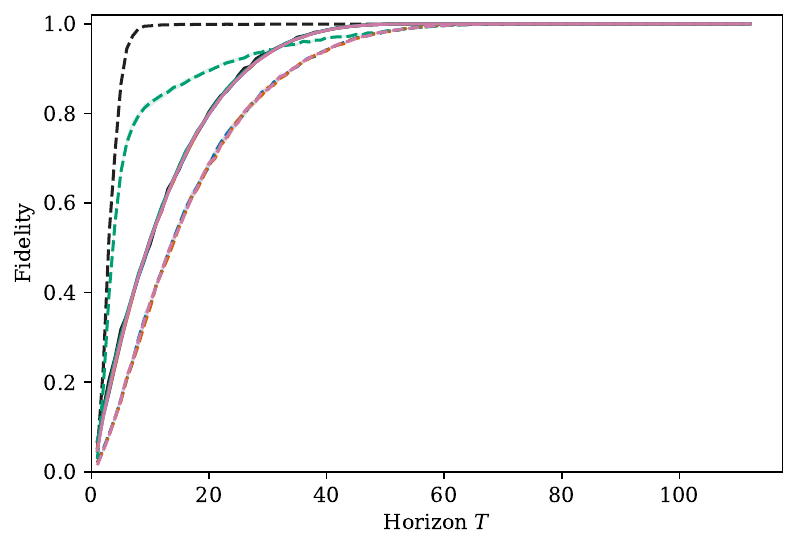}
\end{minipage}
\caption{Matched full-horizon reaching curves for (a)~\(\sect{9}{4}\) (\(C=126\)) and (b)~\(\sect{10}{5}\) (\(C=252\)). Black is the Dicke target; blue, orange, green, and magenta are the phase-uniform, random-sign, random-positive, and Haar ensembles. Solid curves use the generic coupling \((\gamin,\gamsh)=(1,1)\) and dashed curves the coupling \((1.5,1.5)\), at distance \(\varepsilon=0.10\) from full SWAP. Each random-family curve is the median best-over-restart fidelity across 16 paired target vectors, and the shaded band of the same color spans the 25th--75th percentiles across those instances at each horizon. The Dicke target is deterministic and gives a single curve per coupling. The complete horizon range is optimized before the fidelity-dependent hitting times are extracted. At the generic coupling all families reach a given fidelity at nearly the same horizon; at \(\varepsilon=0.10\) the Dicke state is reached much earlier and the random families later.}
\label{fig:target-specificity}
\end{figure*}

The matched full-horizon searches compare the couplings \((\gamin,\gamsh)=(1,1)\) and \((1.5,1.5)\) in \(\sect{9}{4}\) and \(\sect{10}{5}\) for five target families: the Dicke state and four ensembles of targets drawn in the same sector with uniform random phases, random signs, random positive amplitudes, or Haar-random amplitudes (Appendix~\ref{app:targets}). The second coupling lies at distance \(\varepsilon=0.10\) from full SWAP on the ray \(q=1\). This is outside the detunings at which the effective solutions were compared with the microscopic model (\(\varepsilon\leq0.028\) in Sec.~\ref{subsec:eff-vs-micro}), and the one-round remainder \(\|E_\varepsilon-e^{-i\varepsilon H_{\hat{\bm v}}}\|\) evaluated directly is \(0.12\) in \(\sect{9}{4}\) and \(0.16\) in \(\sect{10}{5}\), so Eq.~\eqref{eq:uniform-control-error} gives no control there. The sweep of Sec.~\ref{subsec:crossover} places this coupling at the edge of the slow-time regime: the Dicke horizon there is the slow-time value in \(\sect{8}{4}\) and \(\sect{9}{4}\) and one to two rounds longer in \(\sect{10}{5}\) for \(F\leq0.99\). The study therefore compares a generic coupling with a coupling in the structured region near full SWAP, and the mechanism proposed below is an interpretation. Figure~\ref{fig:target-specificity} shows the curves. At the generic coupling, the curves of the four random families lie close to the Dicke curve, and the observed horizons of all 16 instances of every family agree with the Dicke horizon to within one round at each reported fidelity in both systems, so at this coupling the cost of the tested targets does not depend on their structure. At \(\varepsilon=0.10\) the families separate. The Dicke target reaches each fidelity level earlier than at \((1,1)\), while the phase-uniform, random-sign, and Haar targets reach the same levels later for all 16 paired draws in both systems. Random-positive targets are nearly neutral at \(F=0.95\) and become slower at the higher levels. The random-positive case has a simple consistency check: for amplitudes \(u_s\) drawn uniformly from \([0,1]\), the expected Dicke overlap \(\mathbb E|\langle D_N^{(K)}|\psi_+\rangle|^2\) tends to \((\mathbb Eu)^2/\mathbb Eu^2=3/4\) as \(C\) grows, and random draws at \(C=126\) and \(252\) have a mean Dicke component of \(0.75\). The pattern of Table~\ref{tab:target-deltas}, with no slowdown at \(F=0.95\) and a slowdown that grows toward the Haar value at \(F=0.999\), is consistent with the Dicke component being prepared first and the remainder costing as much as a generic target. A direct test in \(\sect{9}{4}\) supports this reading. For four random-positive targets, with Dicke weights between \(0.72\) and \(0.76\), we record at every horizon the Dicke fidelity \(|\langle D_N^{(K)}|\psi(T)\rangle|^2\) of the optimized final state (Appendix~\ref{app:sweep}). At \(\varepsilon=0.10\) the optimized state first moves close to the Dicke state: at \(T=7\), the horizon at which the Dicke target itself crosses \(F=0.98\), its Dicke fidelity is \(0.90\)--\(0.93\) while the target fidelity is \(0.76\)--\(0.82\), and over the following rounds the Dicke fidelity relaxes to the Dicke weight of the target as the target fidelity grows. At the generic coupling the Dicke fidelity grows together with the target fidelity and approaches the Dicke weight of the target from below.

The exchange graph of the effective theory offers an interpretation. On a positive detuning ray, collect the repeated pulled-back swaps into weights \(w_{ab}\geq0\) and define
\begin{equation}
 \mathcal L_{\hat{\bm v}}=\sum_{a<b}w_{ab}(I-S_{ab}).
 \label{eq:defect-laplacian}
\end{equation}
Every term annihilates the Dicke state, and for a connected exchange graph the Dicke state spans the only zero mode in the sector. The local phase controls preserve the amplitudes of a target but change their relative phases, so target compatibility can be quantified by the smallest graph energy attainable within the local-phase orbit of the target,
\begin{equation}
 E_G(\psi)=\min_{\Theta}
 \langle D(\Theta)\psi|\mathcal L_{\hat{\bm v}}|D(\Theta)\psi\rangle ,
 \label{eq:gauge-frustration}
\end{equation}
where \(D(\Theta)\) is a local phase layer with its global phase removed. On a connected exchange graph, \(E_G=0\) characterizes equal-magnitude targets whose phases lie in the local-phase orbit of the Dicke state. Dicke and local-phase Dicke targets have zero graph energy.

The same compatibility controls the initial steering. From any localized configuration, the phase controls can align every allowed exchange constructively for the Dicke target, and the fidelity derivative reaches the sum of the individual exchange bounds. The sampled phase-uniform and random-sign targets impose competing site-phase conditions and have a smaller optimized derivative. Appendix~\ref{app:target-geometry} derives the bound and reports a steering deficit for every sampled target of these two classes. The same ordering appears in the full-horizon searches. Since the reduction is not controlled at \(\varepsilon=0.10\), and since the graph energy and the initial derivative are first-order quantities that do not by themselves determine a hitting time, this agreement is qualitative.

For each paired target, define \(\Delta\hat T_{\min}(F)=\hat T_{\min}(F;1.5,1.5)-\hat T_{\min}(F;1,1)\). For the phase-uniform, random-sign, and Haar targets, all 16 values of \(\Delta\hat T_{\min}\) are positive at each reported fidelity in both systems, giving the exact two-sided sign-test value \(p=3.05\times10^{-5}\) for each block. The Dicke target has the opposite sign, with the observed horizon at \(F=0.98\) changing from \(22\) to \(7\) rounds in \(\sect{9}{4}\) and from \(38\) to \(8\) in \(\sect{10}{5}\). Appendix~\ref{app:target-geometry} reports the paired medians, including the fidelity-dependent random-positive transition.

The two coupling regimes organize their horizons through different variables,
\begin{equation}
    \begin{cases}
        \dfrac{(N-1)\,\hat T_{\min}(F)}{2(C-1)}\approx\kappa^{(F)},
        & \text{generic},\\[8pt]
        T_{\min}=\dfrac{\tau_*(\hat{\bm v},F)+o(1)}{\varepsilon},
        & \text{near full SWAP}.
    \end{cases}
    \label{eq:two-budget-regimes}
\end{equation}
The generic line is empirical: at the sampled couplings the crossing budget \(\kappa^{(F)}\) is nearly the same in all tested sectors (Fig.~\ref{fig:kappa-collapse}). The near-SWAP line is a theorem at fixed system size, Eq.~\eqref{eq:intrinsic-slow-time}, whose directional coefficient is fixed by the effective control problem and tested against the microscopic dynamics in Fig.~\ref{fig:eff-micro}. On the rays \(q=1\) and \(q=2\) the Dicke horizon passes from the second line to the first between \(\varepsilon=0.1\) and \(0.5\) (Fig.~\ref{fig:crossover} and Table~\ref{tab:sweep}). The crossover and the target ordering of this section are empirical, and the exchange-graph reading is an interpretation.

\section{Discussion}
\label{sec:discussion}

The results distinguish three aspects of state engineering in this collision model: what the control set can reach, how its parametrization is conditioned, and how many rounds a given target costs. Reachability is set by the support structure of the collision map. Singular couplings freeze occupations and confine the trajectory to coordinate sectors, which produces exact fidelity ceilings, while the sampled generic couplings recover the full projective orbit. The same invariant support controls the cost nearby: once the requested fidelity lies above the ceiling, moving a collision angle a small distance away from the singular surface can accumulate the required Fubini--Study distance only over a number of rounds proportional to the inverse displacement. A reachability obstruction is therefore not a binary property of one point of the coupling plane; it shapes the cost in a boundary layer around it.

At the sampled generic couplings the cost follows the control budget: the crossing budget \(\kappa^{(F)}\) is nearly the same in all six sectors and rises smoothly with the requested fidelity, the pre-drawn survey places the bulk of random couplings at the same value, and the matched target study shows the same horizons for generic targets at the generic coupling tested. Near full SWAP the collision sequence reduces to a permutation backbone with weak exchanges. The detuning radius defines the rescaled time, the direction selects the effective control body, and the independently solved effective problem predicts the microscopic horizons; this law is the sharp form of the inverse-distance bound at the full-SWAP corner. The detuning sweep connects the two regimes: on the two tested rays the Dicke horizon follows the slow-time law up to \(\varepsilon=0.1\), Haar targets are delayed in the same range, and beyond \(\varepsilon\approx0.3\) every target returns to the generic cost. In this structured region the target enters through the exchange graph: Dicke amplitudes form its zero mode and local phases align every first-order contribution from a localized input, whereas phase-frustrated targets carry positive graph energy and a steering deficit. The matched searches at \(\varepsilon=0.10\) show the corresponding ordering over the full fidelity range studied. The reduction is not controlled at that detuning, so the exchange-graph mechanism is offered as an interpretation of the observed ordering.

The QFIM adds the local geometry. At the generic coupling the spectrum follows the statistics of a random Gaussian Jacobian, which is the geometric counterpart of a cost that follows parameter counting and ignores the target. Equal endpoint rank can coexist with strongly unequal spectral weights, and a target aligned with the dominant directions can still be prepared quickly in an anisotropic geometry, as the Dicke curves at \(\varepsilon=0.10\) show. A poorly conditioned parametrization in the global sense therefore does not imply a high cost for every target; conditioning and target compatibility enter the cost separately. Across the survey couplings the same holds: the couplings that reach the Dicke target fastest are the least evenly conditioned ones.

The cost in rounds translates into gate counts. One round contains \(L=L_{\rm sh}+L_{\rm in}=K(3n-4)\) partial SWAPs, each of which compiles into three entangling gates in a CNOT/CZ basis~\cite{vu2026data}. For \(\sect{10}{5}\) at \(F=0.98\), the generic crossing at 38 rounds corresponds to \(38\times55=2090\) partial SWAPs and the crossing at \(\varepsilon=0.10\), at 8 rounds, to \(440\). For \(\sect{13}{6}\), Eq.~\eqref{eq:generic-budget} with \(\kappa^{(0.98)}=0.702\) gives about 200 rounds, roughly \(2\times10^4\) partial SWAPs. At generic couplings the cost therefore grows as \(C/(N-1)\), exponentially in \(N\) at fixed filling, whereas explicit circuits prepare \(\Dstate{N}{K}\) with \(O(KN)\) gates~\cite{bartschi2019}, collective measurement with feedback reaches polylogarithmic resources~\cite{yu2024}, and a single local \(Z\) control on an always-on Heisenberg chain reaches \(W\) states and two-excitation Dicke states in times that grow polynomially with \(N\), roughly as \(N^{2.1}\) and \(N^{1.8}\)~\cite{stojanovic2026}. Counted in gates, the present architecture at generic couplings is therefore more expensive than these schemes. What it offers is a different resource profile, two fixed global couplings, local \(\RZ\) rotations as the only time-dependent control, no measurement, and a native excitation-conserving interaction, together with the observation that near full SWAP a structured target is reached at a small fraction of the generic cost. Rounds are also not physical time: a collision at \(\gamma\approx\pi/2\) requires more pulse area than one at \(\gamma=1\), so that in total pulse area \(\sum_\ell|\gamma_{\sigma_\ell}|\) the ratio between the two couplings of Sec.~\ref{subsec:target-selectivity} narrows from \(38/8\) to about \(38/12\) for the Dicke target in \(\sect{10}{5}\), with the ordering unchanged; a complete accounting would report the pulse area per round alongside \(T\). The sweep and the effective model give a first indication: at \(\varepsilon=0.1\) the Dicke target reaches \(F=0.98\) in 7, 7, and 8 rounds for \(C=70\), 126, and 252, while the generic horizon grows from 14 to 38 rounds, and the slow-time value \(\tau_*\) at \(F=0.98\) stays between \(0.64\) and \(0.73\) for \(C=35\), 70, and 126 on three rays and is \(0.655\) for \(C=252\) on the ray \(q=1\). Whether \(\tau_*(N,K)\) for the Dicke target grows only polynomially with \(N\) while the generic cost grows as \(C/(N-1)\) is, in our view, the central open question raised by these results; it can be answered within the effective model without microscopic optimization.

Several limitations should be kept in mind, and the results have different standing. All results concern the fixed-excitation sector reachable from the \(K\)-excitation localized input of this architecture, so states with indefinite excitation number, such as GHZ states, are outside the control set. The invariant-sector ceilings, the inverse-distance bound, and the slow-time theorem are proved, the last at fixed system size under the regularity condition of Appendix~\ref{app:slow-time}, with its microscopic validation limited to \(N\leq8\) and \(\varepsilon\leq0.028\), and for the Dicke target in \(\sect{8}{4}\) and \(\sect{9}{4}\) on two rays to \(\varepsilon\leq0.1\) at \(F\leq0.99\). The projective-orbit certificates are numerical and cover \(C\leq21\); the generator argument of Appendix~\ref{app:lie-proof} is a sketch with a stated gap. The control-budget organization is empirical, rests on four couplings in systems with \(N\leq13\), and is an upper envelope: the pre-drawn survey contains weak-collision couplings, with \(\gamsh\) close to \(0\) or \(\pi\), that are two to three times faster and are not analyzed here. The target study covers two sectors and one coupling pair, and the detuning sweep two rays, two sectors, and Dicke and Haar targets, with half the restarts of the main campaigns; all optimizations are initialized near the passive schedule. The crossover and the target ordering are empirical, and the exchange-graph reading is an interpretation. The QFIM statistics rest on small fixed banks of phase vectors. Observed horizons are optimizer-dependent upper bounds on the intrinsic minimum. Platforms with a pure \(XY\) interaction are not covered. The scaling of \(\tau_*\) with \(N\) and \(K\) beyond \(C=252\), the crossover along other rays and in larger sectors, the weak-collision corners, controllability inside the invariant sectors, and a closed-form predictor for \(\tau_*(N,K,\hat{\bm v},F)\) remain open.

The phase layers need not be applied after every round. A binary schedule \(\chi_t\in\{0,1\}\) with \(c_{t+1}=\RZ(\chi_t\theta_t)Mc_t\) contains the passive sequence \(M^T\) of Ref.~\cite{vu2026pra}, an initial passive block followed by controlled rounds, and periodic phase insertion as special cases, and the control density \(\rho_R=T^{-1}\sum_t\chi_t\) parametrizes the cost of sparse or delayed control; near full SWAP this leads to a schedule-dependent coefficient \(\tau_*(\hat{\bm v},F;\{\chi_t\})\). A second direction concerns learned controllers. Each control instance in this paper is optimized independently; a model trained across \((N,K,\gam)\) or across targets, of the kind our recent work~\cite{vu2026sd} explores in the minimal-control limit of two global angles, would have to represent the gauge-reduced control count, the exchange graph of a detuning ray, the invariant supports, and the target compatibility, and would remain constrained by the reachability, conditioning, and target-compatibility structures identified here.

\begin{acknowledgments}
\"O.E.M. and F.O. acknowledge financial support from the Scientific and Technological Research Council of T\"urkiye (T\"UB\.ITAK), Project Grant No.~126F097. F.O. acknowledges financial support from the Tokyo International University Personal Research Fund and Special Grant-in-Aid for Research Work. AI-based tools assisted with drafting, editing, and checking the manuscript; the authors reviewed all content, numbers, and references and take full responsibility for the manuscript.
\end{acknowledgments}

\section*{Data availability}

The code, the coupling lists, the optimizer settings, the target vectors, the data tables behind every figure and quoted number, and the figure scripts are maintained at \url{https://github.com/KT-One/Multi-Rz}. The detuning sweep of Fig.~\ref{fig:crossover} and the effective curves for \(\sect{9}{4}\) and \(\sect{10}{5}\) were computed with the same code, and their scripts and data will be added to the repository. The per-restart records of all optimization campaigns are retained by the authors and will be deposited together with a tagged version of the code in a versioned public archive.

\appendix

\section{Numerical methods}
\label{app:methods}

\subsection{Numerical conventions}
State amplitudes use double precision (\texttt{complex128}). At a sampled horizon \(T\), \(F_{\rm best}(T)\) is the best fidelity within the restart ensemble of that study, and Eq.~\eqref{eq:observed-hitting} defines \(\hat T_{\min}(F)\) on the horizon set searched in that study. The hard rank of the QFIM uses the relative eigenvalue cutoff \(10^{-9}\lambda_{\max}\). The QFIM evaluations use stored phase vectors. Figure~\ref{fig:effective-rank-landscape} uses a bank of four vectors with entries drawn independently and uniformly from \([-1,1]\) (seeds \(20260904\)--\(20260907\), recorded with the data); the displayed landscape is the first vector, and the robustness check reports the spread of the participation ratio over the four. Figure~\ref{fig:synthesis} evaluates, at every horizon, the passive schedule \(\Theta=0\) and four draws with independent Gaussian entries of standard deviation \(0.25\), the initialization distribution of the optimizations; the plotted curves average over the four Gaussian vectors. Since the lower edge of the Marchenko--Pastur spectrum approaches zero at \(\kappa=1\), the hard rank near \(\kappa=1\) is sensitive to the cutoff; the eigenvalue gap at the cutoff is not reported here. The un-hatted \(T_{\min}\) of Appendix~\ref{app:slow-time} is the intrinsic optimal-control hitting time.

The generic campaign of Fig.~\ref{fig:kappa-collapse} contains the six sectors \(\sect{8}{4}\) (\(C=70\)), \(\sect{9}{4}\) (\(126\)), \(\sect{10}{5}\) (\(252\)), \(\sect{11}{5}\) (\(462\)), \(\sect{12}{6}\) (\(924\)), and \(\sect{13}{6}\) (\(1716\)). Every sector uses the same four coupling pairs, \((\gamin,\gamsh)=(1.1,0.7)\), \((1.3,0.9)\), \((1,1)\), and \((0.8,1.2)\), which gives 24 curves. Each sampled horizon uses eight cold restarts, one at zero and seven Gaussian with \(\sigma=0.25\), followed by 2000 Adam epochs and 100 L-BFGS steps. The stored horizon curves are converted to monotone best-fidelity envelopes before \(\kappa^{(F)}\) is evaluated.

The finite-radius scaling survey of Sec.~\ref{subsec:predictive} uses four seeds with 256 batched Adam restarts, 1500 epochs, learning rate \(0.02\), and patience 250. Its \(F=0.98\) crossings provide the \(1/\varepsilon\) scaling check. The effective-to-microscopic comparison uses the separate protocol of Appendix~\ref{app:eff-micro}.

The target study of Fig.~\ref{fig:target-specificity} uses one protocol for every target family and both couplings. All integer horizons through 63 in \(\sect{9}{4}\) and 112 in \(\sect{10}{5}\) are evaluated with eight cold restarts and 2000 Adam epochs at learning rate \(0.02\). The complete horizon range is optimized before any fidelity-dependent hitting time is extracted. The same target vector is paired across the two couplings, and the displayed median and interquartile band are computed across targets after taking the best restart at each horizon.

\subsection{Target ensembles}
\label{app:targets}
Every target lies in the \(C\)-dimensional sector. Besides the Dicke state, Fig.~\ref{fig:target-specificity} uses the ensembles
\begin{equation}
\begin{aligned}
 \ket{\psi_{\rm phase}}
 &=\frac1{\sqrt C}\sum_s e^{i\varphi_s}\ket{s},\\
 \ket{\psi_{\pm}}
 &=\frac1{\sqrt C}\sum_s\sigma_s\ket{s},\\
 \ket{\psi_+}
 &=\frac{\sum_s u_s\ket{s}}{(\sum_su_s^2)^{1/2}},
\end{aligned}
\label{eq:target-ensembles}
\end{equation}
with \(\varphi_s\sim\mathrm{Unif}(0,2\pi)\), \(\Pr(\sigma_s=\pm1)=1/2\), and \(u_s\sim\mathrm{Unif}(0,1)\), all independent within each draw. Haar targets are obtained by normalizing a vector with independent complex Gaussian components in the sector. Each random ensemble contains 16 paired draws per system, and the same target vector is used at both couplings, so every reported horizon difference is paired target by target.

\subsection{Pre-drawn coupling survey}
\label{app:generic-sampling}
The independent survey draws 36 pairs uniformly from \([0,\pi]^2\), rejecting a pair when either coordinate lies within \(0.15\) of \(0\), \(\pi/2\), or \(\pi\). The seed is \(20260901\), and the complete list is stored before optimization. The same pairs are used in both sectors, and all 72 results are retained. Table~\ref{tab:generic-survey} reports the distribution of the observed integer-horizon budgets at \(F=0.98\).

\begin{table}[!tbp]
\caption{Pre-drawn coupling survey at \(F=0.98\). Each sector contains 36 coupling pairs drawn before optimization, and every search crosses this level. The last column is the spacing of the allowed \(\kappa\) values for integer horizons.}
\label{tab:generic-survey}
\begin{ruledtabular}
\begin{tabular}{lrrrr}
Sector & Median & Minimum & Maximum & Step\\
\colrule
\sect{7}{3} & $0.706$ & $0.353$ & $0.706$ & $0.088$\\
\sect{8}{4} & $0.659$ & $0.254$ & $0.710$ & $0.051$\\
\end{tabular}
\end{ruledtabular}
\end{table}

\subsection{Detuning sweep}
\label{app:sweep}
The sweep uses the couplings \(\gamsh=\pi/2-\varepsilon v_{\rm sh}\) and \(\gamin=\pi/2-\varepsilon v_{\rm in}\) on the rays \((v_{\rm sh},v_{\rm in})=(1,1)/\sqrt2\) and \((1,2)/\sqrt5\), that is \(q=1\) and \(q=2\), with \(\varepsilon\in\{0.05,0.1,0.2,0.3,0.5,0.8\}\), the Dicke target, and four Haar-random targets per sector, drawn as in Appendix~\ref{app:targets}, the same on both rays, and different from those of Fig.~\ref{fig:target-specificity}. The optimizer is the full-horizon recipe of Fig.~\ref{fig:target-specificity}: Adam with 2000 epochs and learning rate \(0.02\), one restart from \(\Theta=0\) and the others Gaussian with \(\sigma=0.25\), here with two seed banks of eight restarts, half of the restarts used there. Every integer horizon is optimized up to the cap \(\lceil\max\{1.6/\varepsilon,\,1.3\cdot2(C-1)/(N-1)\}\rceil\), which is 26 rounds in \(\sect{8}{4}\) and 41 in \(\sect{9}{4}\); at \(\varepsilon=0.05\) the scans were extended to 80 rounds. Where the sweep overlaps the earlier campaigns it reproduces them: at \(\varepsilon=0.8\), where the couplings are \((1.005,1.005)\) and \((0.855,1.213)\), all targets cross \(F=0.98\) at 14 rounds in \(\sect{8}{4}\) and at 21 or 22 in \(\sect{9}{4}\), the generic values of Fig.~\ref{fig:kappa-collapse}, and at \(\varepsilon=0.1\) in \(\sect{9}{4}\) the Dicke target crosses at 7 rounds and the Haar targets at 28--30, against 7 and 29--31 in Fig.~\ref{fig:target-specificity}. Table~\ref{tab:sweep} lists the observed horizons. The effective curves for \(\sect{9}{4}\) on the rays \(q=0.5\), 1, and 2 use the effective solver of Sec.~\ref{subsec:eff-vs-micro} unchanged, with the settings of the other sectors (8 seeds, 600 epochs, learning rate \(0.05\)) on a slightly finer grid of \(\tau_{\rm sh}\), and without access to microscopic data; the curve for \(\sect{10}{5}\) on the ray \(q=1\) uses four seeds, whose best values differ by less than \(3\times10^{-3}\) in fidelity at every grid point. The trajectory test of Sec.~\ref{subsec:target-selectivity} uses the optimizer of the sweep at the couplings \((1,1)\) and \((1.5,1.5)\) in \(\sect{9}{4}\), every integer horizon up to 41, and four random-positive targets drawn as in Appendix~\ref{app:targets}; at each horizon it records the target fidelity and the Dicke fidelity of the final state of the best restart.

\begin{table}[!tbp]
\caption{Observed horizons \(\hat T_{\min}(F)\) of the detuning sweep on the rays \(q=1\) and \(q=2\). Haar entries give the range over four targets; \(>T\) means that the level was not reached within the cap \(T\) of the scan.}
\label{tab:sweep}
\centering
\scriptsize
\setlength{\tabcolsep}{2.0pt}
\begin{tabular}{lcrrrrrrrr}
\toprule
 & & \multicolumn{4}{c}{Dicke} & \multicolumn{4}{c}{Haar} \\
\cmidrule(lr){3-6}\cmidrule(lr){7-10}
Sector & $\varepsilon$ & $.95$ & $.98$ & $.99$ & $.999$ & $.95$ & $.98$ & $.99$ & $.999$ \\
\midrule
\multicolumn{10}{l}{$q=1$}\\
\sect{8}{4} & 0.05 & 12 & 13 & 14 & 17 & 32--35 & 38--41 & 43--44 & 51--52 \\
 & 0.1 & 6 & 7 & 7 & 10 & 18--19 & 20--22 & 23 & $>26$ \\
 & 0.2 & 5 & 7 & 9 & 14 & 12--13 & 15 & 15--17 & 19 \\
 & 0.3 & 9 & 12 & 13 & 18 & 12 & 14 & 15--16 & 18 \\
 & 0.5 & 12 & 14 & 15 & 18 & 12 & 14 & 15 & 18 \\
 & 0.8 & 12 & 14 & 15 & 17 & 12 & 14 & 15 & 18 \\
\addlinespace[2pt]
\sect{9}{4} & 0.05 & 12 & 13 & 14 & 16 & 43--45 & 52--54 & 56--58 & 69--72 \\
 & 0.1 & 6 & 7 & 8 & 12 & 25 & 28--30 & 31--33 & 37--38 \\
 & 0.2 & 8 & 13 & 15 & 27 & 19 & 22--23 & 25 & 29--30 \\
 & 0.3 & 15 & 21 & 23 & 28 & 18--19 & 22 & 24 & 28--29 \\
 & 0.5 & 18 & 22 & 24 & 27 & 18--19 & 22 & 23--24 & 28 \\
 & 0.8 & 18 & 22 & 24 & 28 & 18--19 & 22 & 24 & 28 \\
\midrule
\multicolumn{10}{l}{$q=2$}\\
\sect{8}{4} & 0.05 & 11 & 12 & 13 & 16 & 35--37 & 42--44 & 44--46 & 52--56 \\
 & 0.1 & 6 & 7 & 7 & 9 & 18--20 & 21--23 & 24--25 & $>26$ \\
 & 0.2 & 4 & 6 & 8 & 17 & 12--13 & 15--16 & 16--17 & 19--20 \\
 & 0.3 & 9 & 12 & 14 & 18 & 11--12 & 14 & 15--16 & 18 \\
 & 0.5 & 12 & 14 & 15 & 18 & 12 & 14 & 15 & 18 \\
 & 0.8 & 12 & 14 & 15 & 18 & 12 & 14 & 15 & 18 \\
\addlinespace[2pt]
\sect{9}{4} & 0.05 & 11 & 13 & 13 & 15 & 43--46 & 53--54 & 56--59 & 68--71 \\
 & 0.1 & 6 & 7 & 8 & 13 & 25--26 & 28--30 & 30--32 & 38--39 \\
 & 0.2 & 8 & 11 & 14 & 25 & 19 & 22--23 & 24--25 & 29 \\
 & 0.3 & 14 & 20 & 23 & 28 & 18--19 & 22 & 24 & 28 \\
 & 0.5 & 19 & 22 & 24 & 28 & 18--19 & 22 & 23--24 & 28 \\
 & 0.8 & 18 & 21 & 24 & 28 & 18--19 & 22 & 23--24 & 28 \\
\bottomrule
\end{tabular}
\end{table}

\section{Collision invariants}
\label{app:singular-surveys}

\subsection{Collision sequence}
\label{app:schedule}
Set \(n=N-K\), \(R=\lfloor n/2\rfloor\), and \(S=\lceil n/2\rceil\). With zero-based site indices, the shuttle sites are \(A_\alpha=\alpha\), the first chain is \(r_i=K+i\), and the second is \(s_i=K+R+i\). A round visits \(\alpha=0,\ldots,K-1\) in order. For each shuttle, \(i=0,\ldots,S-1\) is visited in order, with the following interactions executed chronologically, a line being omitted when its condition fails,
\begin{equation}
\begin{array}{ll}
 (A_\alpha,r_i;\mathrm{sh}) & i<R,\\
 (r_i,r_{i-1};\mathrm{in}) & 0<i<R,\\
 (r_i,r_{i+1};\mathrm{in}) & i<R-1,\\
 (A_\alpha,s_i;\mathrm{sh}) & i<S,\\
 (s_i,s_{i-1};\mathrm{in}) & 0<i<S,\\
 (s_i,s_{i+1};\mathrm{in}) & i<S-1.
\end{array}
\label{eq:complete-schedule}
\end{equation}
The triple specifies the two sites and the collision family. An internal chain edge may be visited twice during a subround. For \(n\geq2\), a full round contains \(L_{\rm sh}=Kn\) shuttle collisions and \(L_{\rm in}=2K(n-2)\) intra-register collisions. Equation~\eqref{eq:ordered-collision-word} uses exactly this sequence with later factors to the left.

\subsection{Invariant support}
\label{app:component-proof}
The local phase torus distinguishes every computational-basis \(K\)-subset. If a subspace is invariant under \(D(\theta)M\) for every \(\theta\), it is invariant under \(M\) and, by unitarity, under every \(D(\theta)\), and must therefore be a coordinate subspace. The support components of \(M\) consequently bound the controlled reachable support.

On \(\gamin=\pi/2\), the prefix construction of Eq.~\eqref{eq:defect-prefixes} pulls every nontrivial shuttle rotation back to a transposition of logical labels. Following the sequence of Eq.~\eqref{eq:complete-schedule} shows that, when \(n\geq4\), these transpositions connect \(N-2\) labels and leave two labels fixed,
\begin{equation}
 f_R=r_{R-1},\qquad f_S=s_{S-1}.
\end{equation}
Both the transpositions and the permutation backbone fix these two labels. Their occupations \(a,b\in\{0,1\}\) are conserved by every round and by every local phase layer, giving the sectors
\begin{align}
 \Omega_{ab}&=\{s:|s|=K,\ n_{f_R}(s)=a,\ n_{f_S}(s)=b\},
 \label{eq:occupation-sectors}\\
 |\Omega_{ab}|&=\binom{N-2}{K-a-b},
 \label{eq:component-binomial}
\end{align}
where a binomial coefficient is zero when its lower argument is outside its range. At \(n=3\) the construction leaves one isolated label, \(s_1\), and the two sectors have sizes \(\binom{N-1}{K-a}\). At \(n\leq2\) the intra-register family is absent.

For each fixed \((N,K)\), these invariant sectors are the support components at every shuttle angle outside a finite exceptional set. Expand the one-round amplitudes about full SWAP. Every pulled-back exchange edge has a nonzero first-order coefficient because repeated occurrences add positive multiplicity, and the backbone transitions have nonzero leading coefficients. Choose a spanning tree in each group orbit generated by these exchanges and the backbone. Each selected transition amplitude is a nonzero trigonometric polynomial with finitely many zeros in \([0,\pi]\), and outside their finite union the spanning trees remain in the support graph. The exceptional set may include angles outside \(\Gamma_{\rm s}\).

For a normalized state supported on \(\Omega\) and a target \(\ket{\psi_{\rm tar}}\), the Cauchy--Schwarz inequality gives \(|\langle\psi_{\rm tar}|c\rangle|^2\leq\sum_{s\in\Omega}|\langle s|\psi_{\rm tar}\rangle|^2\), the weight of the target on \(\Omega\). For the Dicke target,
\begin{equation}
 F_D=\frac1C\left|\sum_{s\in\Omega}c_s\right|^2
 \leq\frac{|\Omega|}{C}.
 \label{eq:support-cauchy}
\end{equation}
The localized input has zero excitation on the isolated chain labels, so substituting \(a=0\) or \(a=b=0\) gives the ceilings of Eq.~\eqref{eq:fmax-comp}. For \(\sect{6}{2}\), the four sectors have sizes \(6,4,4,1\). The support argument proves the ceiling, and the finite-system optimizations establish attainment in the sampled sectors.

\subsection{Projective certificates}
Tables~\ref{tab:appA} and \ref{tab:appB} summarize the finite-system projective-orbit certificates and the matching support ceilings. The sampled generic couplings attain the full certificate dimension \(d_{\rm orb}=2(C-1)\). The frozen-shuttle lines and the permutation points, with both angles in \(\Gamma_{\rm s}\), remain on the initial ray. The \(\gamin=\pi/2\) line occupies the invariant sectors derived above.

The certificate is computed by constructing the drift-conjugated algebra \(\mathfrak h(\gam)\) of Eq.~\eqref{eq:lie} and projecting \(\{X\ket\psi:X\in\mathfrak h(\gam)\}\) off the state; the resulting \(d_{\rm orb}\) is the projective-orbit certificate. Direct control optimization at \(\kappa=3\) evaluates \(F_{\rm best}\) for the same sectors. On \(\gamin=\pi/2\), the localized input has the analytic ceiling of Eq.~\eqref{eq:fmax-comp}, and the survey reaches these ceilings in the tested sectors.

\begin{table}[!tbp]
\caption{State-orbit certificates and invariant-support dimensions across representative sectors and couplings. For generic couplings, full projective-orbit certification gives \(d_{\rm orb}=2(C-1)\). On \(\gamin=\pi/2\), the entries are the invariant-support projective dimensions \(2(|\Omega|-1)\). The frozen-shuttle and corner entries remain on the initial ray. ``Corner'' denotes both angles in \(\Gamma_{\rm s}\).}
\label{tab:appA}
\begin{ruledtabular}
\begin{tabular}{lrrrrr}
Coupling & \sect{4}{1} & \sect{5}{2} & \sect{6}{2} & \sect{6}{3} & \sect{7}{2} \\
\colrule
$C$ & 4 & 10 & 15 & 20 & 21 \\
$2(C-1)$ & 6 & 18 & 28 & 38 & 40 \\
\colrule
Generic & 6 & 18 & 28 & 38 & 40 \\
$\gamsh{=}\pi/2$ & 6 & 18 & 28 & 38 & 40 \\
$\gamin{\in}\{0,\pi\}$ & 6 & 18 & 28 & 38 & 40 \\
$\gamin{=}\pi/2$ & 4 & 10 & 10 & 18 & 18 \\
$\gamsh{\in}\{0,\pi\}$ & 0 & 0 & 0 & 0 & 0 \\
Corner & 0 & 0 & 0 & 0 & 0 \\
\end{tabular}
\end{ruledtabular}
\end{table}

\begin{table}[!tbp]
\caption{Best attained Dicke fidelity \(F_{\rm best}\) for the localized input. Generic couplings and the full-SWAP shuttle line with a nonsingular \(\gamin\) attain unit fidelity in these sectors. The \(\gamin=\pi/2\) row attains the analytic invariant-support ceiling \(|\Omega|/C\). Frozen and monomial couplings have singleton output support and give the exact value \(1/C\).}
\label{tab:appB}
\begin{ruledtabular}
\begin{tabular}{lrrr}
Coupling & \sect{5}{2} & \sect{6}{3} & \sect{7}{3} \\
\colrule
Generic / $\gamsh{=}\pi/2$ & $1.000$ & $1.000$ & $1.000$ \\
$\gamin{=}\pi/2$ & $0.600$ & $0.500$ & $0.286$ \\
$\gamsh{\in}\{0,\pi\}$ & $0.100$ & $0.050$ & $0.029$ \\
\end{tabular}
\end{ruledtabular}
\end{table}

\section{Projective generators (sketch)}
\label{app:lie-proof}
This appendix is a proof sketch; the step that it leaves open is stated at the end, and the reachability statements of the main text rest on the computed certificates of Appendix~\ref{app:singular-surveys}. The projective-algebra argument isolates the exchange edges without requiring the one-body and pair-diagonal occupations to span the full traceless diagonal algebra. Let
\begin{equation}
 D(u,w)=\sum_i u_i n_i+\sum_{i<j}w_{ij}n_i n_j .
\end{equation}
On a token edge \(R\cup\{a\}\leftrightarrow R\cup\{b\}\), its diagonal gap is
\begin{equation}
 \delta_{a,b;R}(D)=u_a-u_b+\sum_{j\in R}(w_{aj}-w_{bj}).
 \label{eq:quadratic-gap}
\end{equation}
Distinct token edges define distinct gap functionals up to sign, so a generic quadratic diagonal has nonzero, pairwise distinct absolute gaps. If \(X_e\) is the symmetric edge quadrature,
\begin{equation}
 \operatorname{ad}_{iD}^2(iX_e)=-\delta_e^2 iX_e .
\end{equation}
Finite Vandermonde interpolation isolates every \(iX_e\); one commutator produces \(iY_e\), and connectivity of the token graph generates the full projective algebra. Centering the diagonal generator, \(\widetilde D=D-\operatorname{tr}(D)I/C\), gives
\begin{equation}
 \Lie(i\widetilde D,\{iX_e,iY_e\})=\mathfrak{su}(C).
 \label{eq:projective-suC}
\end{equation}
The scalar center removed by this centering changes no projective dynamics.

The step that is not established is that \(\mathfrak h(\gam)\) contains a generic quadratic diagonal. The controls themselves generate only one-body diagonals \(D(u,0)=\sum_iu_in_i\), for which the gap of every token edge \(R\cup\{a\}\leftrightarrow R\cup\{b\}\) equals \(u_a-u_b\) independently of \(R\) when \(K\geq2\); edges that share the pair \(\{a,b\}\) are then degenerate under \(\operatorname{ad}^2_{iD}\), and the Vandermonde step cannot separate them. Pair-diagonal terms \(w_{ij}\neq0\) would have to arise from commutators of the drift-conjugated generators, which we have not shown in general. The certificates of Appendix~\ref{app:singular-surveys} compute \(d_{\rm orb}\) directly and do not depend on this step. Subspace-controllability techniques for Heisenberg chains with limited local control~\cite{wang2016} are a possible route to a complete argument.

\section{Singular cost bound}
\label{app:singular-speed}
Let \(\gamma_\alpha^\star\) be a collision angle at which every controlled endpoint has target fidelity at most \(F_0\). Freeze an arbitrary control sequence and vary only that angle. For \(U(\gamma_\alpha)=e^{i\gamma_\alpha S}\) with \(\|S\|=1\), differentiating one of the \(T L_\alpha\) occurrences changes the endpoint at Fubini--Study speed at most one. Summing the occurrences gives
\begin{equation}
 d_{\rm FS}(\psi_T(\gamma_\alpha),\psi_T(\gamma_\alpha^\star))
 \leq T L_\alpha|\gamma_\alpha-\gamma_\alpha^\star| .
 \label{eq:fs-parameter-speed}
\end{equation}
Every singular endpoint is at target angle at least \(\arccos\sqrt{F_0}\), whereas fidelity \(F\) places the perturbed endpoint within \(\arccos\sqrt F\). The triangle inequality then gives Eq.~\eqref{eq:singular-boundary-speedlimit}. If both families are detuned at once, by \(\varepsilon\hat v_{\rm in}\) and \(\varepsilon\hat v_{\rm sh}\), the same summation over the \(TL\) occurrences gives \(d_{\rm FS}\leq T\varepsilon\Lambda\) with \(\Lambda=\sum_\ell|\hat v_{\sigma_\ell}|\), and the ceiling \(F_0=1/C\) at the permutation point \((\pi/2,\pi/2)\) yields Eq.~\eqref{eq:corner-bound} after multiplication by \(\varepsilon\). The collision counts \(L_{\rm in}=2K(N-K-2)\) and \(L_{\rm sh}=K(N-K)\) follow from Eq.~\eqref{eq:complete-schedule}; deterministic checks reproduce both counts in all 840 main-regime sectors with \(N\leq60\).

\section{Near-SWAP control}
\label{app:slow-time}

\subsection{Controlled approximation}
Let \(R_t=D_t\Pi\) and \(W_t=R_{t-1}\cdots R_0\). Repeatedly moving the factors \(R_t\) to the left gives
\begin{equation}
 U^{(T)}=i^{LT}W_T E_{T-1}\cdots E_0,\qquad
 E_t=W_t^{-1}E_\varepsilon W_t .
\label{eq:exact-controlled-frame}
\end{equation}
Write \(A_\ell=\hat v_{\sigma_\ell}\widetilde S_\ell\). Since \(\|A_\ell\|=|\hat v_{\sigma_\ell}|\), the unitary product remainder satisfies
\begin{equation}
 \left\|\overleftarrow{\prod_\ell}e^{-i\varepsilon A_\ell}
       -e^{-i\varepsilon\sum_\ell A_\ell}\right\|
 \leq\varepsilon^2\Lambda^2 .
 \label{eq:one-round-error}
\end{equation}
Unitary conjugation leaves this estimate unchanged. Telescoping the \(T\) factors gives the linear-in-\(T\) bound of Eq.~\eqref{eq:uniform-control-error}. For \(\varepsilon T\leq\tau_{\max}\), the total error is at most \(\tau_{\max}\varepsilon\Lambda^2\), uniformly over detuning-dependent phase sequences.

\subsection{Effective controls}
A product of phase layers and permutations has the form \(W_t=D(\Theta_t)\Pi^t\). The cumulative phases obey \(\Theta_{t+1}=\theta_t+\Pi\cdot\Theta_t\), where \(\Pi\cdot\Theta_t\) denotes the relabeling of the entries of \(\Theta_t\) induced by the permutation \(\Pi\). Apart from the fixed initial frame, the cumulative phases can be chosen independently. If \(q=\ord(\Pi)\), the generator available at residue \(r=t\bmod q\) belongs to
\begin{equation}
 \mathcal O_r=
 \{\Pi^{-r}D(-\Theta)H_{\hat{\bm v}}D(\Theta)\Pi^r:\Theta\}.
 \label{eq:residue-orbits}
\end{equation}
Every block of \(q\) rounds visits each residue once. Its average generator lies in a Minkowski average of these sets, and rapid switching gives the compact convex effective set
\begin{equation}
 \mathcal K_{\hat{\bm v}}
 =\conv\!\left(\frac1q\sum_{r=0}^{q-1}\mathcal O_r\right)
 =\frac1q\sum_{r=0}^{q-1}\conv(\mathcal O_r),
 \label{eq:effective-control-set}
\end{equation}
where the sum is a Minkowski sum that preserves the residue frequencies fixed by the microscopic sequence. The effective evolution is
\begin{equation}
 \frac{dV}{d\tau}=-iG(\tau)V,
 \qquad G(\tau)\in\mathcal K_{\hat{\bm v}},\quad V(0)=I .
 \label{eq:effective-control-evolution}
\end{equation}
For fixed \((N,K)\) and a fixed ray, the per-block remainder is quadratic in \(q\varepsilon\), so over a bounded interval of rescaled time it vanishes with \(\varepsilon\). Limits of the block averages give admissible effective controls. Conversely, a piecewise-constant effective control can be approximated by finite mixtures of the residue-constrained block generators, then by actual phase sequences. Together with the uniform product bound, these two constructions give convergence of the reachable value functions on compact intervals of rescaled time.

\subsection{Directional geometry}
\label{app:directional-body}
The ray direction enters the effective body through the exchange weights. For residue \(r\), write the exchange part of an admissible generator as
\begin{equation}
\begin{aligned}
 h_r(\Theta;\hat{\bm v})
 &=\Delta_r(\hat{\bm v})
 +\sum_e c_{r,e}(\hat{\bm v})
 \bigl[\cos\varphi_e X_e+\sin\varphi_eY_e\bigr],\\
 c_{r,e}(\hat{\bm v})
 &=m^{\rm in}_{r,e}\hat v_{\rm in}
 +m^{\rm sh}_{r,e}\hat v_{\rm sh}.
\end{aligned}
\label{eq:directional-edge-coefficients}
\end{equation}
The integers \(m^{\rm in}_{r,e}\) and \(m^{\rm sh}_{r,e}\) count the pulled-back full-SWAP transpositions contributing to edge \(e\). The phase difference on that edge aligns either sign of its \(X_e\) projection. Since support functions add under Minkowski sums, the corresponding half-width of \(\mathcal K_{\hat{\bm v}}\) is
\begin{equation}
 \rho_e(\hat{\bm v})
 =\frac1q\sum_{r=0}^{q-1}
 \left|c_{r,e}(\hat{\bm v})\right|,
 \label{eq:directional-edge-width}
\end{equation}
and we normalize the profile by
\begin{equation}
 \widehat\rho_e(\hat{\bm v})
 =\frac{\rho_e(\hat{\bm v})}{\sum_f\rho_f(\hat{\bm v})}.
 \label{eq:normalized-edge-width}
\end{equation}
Exact transposition counts give distinct profiles for the rays \(\hat{\bm v}\propto(1,2)\) and \(\hat{\bm v}\propto(2,1)\). In \(\sect{6}{3}\), the edge \(e=\{4,5\}\) satisfies
\begin{equation}
 \left|
 \widehat\rho_e(1,2)-\widehat\rho_e(2,1)
 \right|=\frac{9}{56},
 \label{eq:width-certificate-d63}
\end{equation}
and in \(\sect{7}{3}\) the edge \(e=\{0,1\}\) gives
\begin{equation}
 \left|
 \widehat\rho_e(1,2)-\widehat\rho_e(2,1)
 \right|=\frac1{30}.
 \label{eq:width-certificate-d73}
\end{equation}
The common Euclidean normalization of each ray cancels in \(\widehat\rho_e\), and these exact differences certify distinct effective-body shapes for the two directions.

The same chronology gives a global positive-ray statement. Parametrize positive rays by the ratio \(t=v_{\rm in}/v_{\rm sh}\). The pulled-back intra-register and shuttle width vectors are linearly independent throughout \(N\geq5\) and \(2\leq K\leq\lfloor N/2\rfloor\), therefore
\begin{equation}
 t\neq t'\quad\Longrightarrow\quad\widehat\rho(t)\neq\widehat\rho(t').
 \label{eq:positive-ray-anisotropy}
\end{equation}
Distinct positive detuning rays consequently produce effective control bodies that are not homothetic. The deterministic census through \(N\leq60\) checks the exact collision bookkeeping in 840 sectors and finds no failed witness.

\subsection{Minimum time}
The final local phase layer remains available at no cost in rescaled time. Define
\begin{equation}
 \Phi(\psi)=\max_\Theta\bigl|\bra{D_N^{(K)}}D(\Theta)\ket{\psi}\bigr|^2 .
 \label{eq:phase-terminal-payoff}
\end{equation}
Permutation symmetry leaves the Dicke payoff invariant under the final permutation. Let \(\mathcal F_{\rm eff}(\tau)\) be the supremum of \(\Phi\) reachable by effective time at most \(\tau\), and let the discrete value optimize over all legal controls and all integer horizons with \(\varepsilon T\leq\tau\). The preceding construction gives uniform convergence of these values on bounded intervals. Set
\begin{align}
 \tau_{\geq}(F)&=\inf\{\tau:\mathcal F_{\rm eff}(\tau)\geq F\},\\
 \tau_{>}(F)&=\inf\{\tau:\mathcal F_{\rm eff}(\tau)>F\}.
\end{align}
Uniform value convergence implies
\begin{equation}
\begin{aligned}
 \tau_{\geq}(F)
 &\leq\liminf_{\varepsilon\to0}\varepsilon T_{\min}(\varepsilon;F)\\
 &\leq\limsup_{\varepsilon\to0}\varepsilon T_{\min}(\varepsilon;F)
 \leq\tau_{>}(F).
\end{aligned}
\label{eq:hitting-time-sandwich}
\end{equation}
A regular level has \(\tau_{\geq}(F)=\tau_{>}(F)<\infty\). Their common value is \(\tau_*\), which yields Eq.~\eqref{eq:intrinsic-slow-time} for the intrinsic optimum at fixed system size. The deterministic validation of the approximation contains 240 propagation comparisons organized into 48 refinement sequences with decreasing errors.

\subsection{Endpoint bound}
Let \(\mathcal P=n_{f_S}\) project onto the protected endpoint occupation and \(\mathcal Q=I-\mathcal P\). The pulled-back chronology contains exactly \(2K\) intra-register occurrences crossing this cut and no shuttle occurrence, so every \(G\in\mathcal K_{\hat{\bm v}}\) obeys
\begin{equation}
 \|\mathcal PG\mathcal Q\|\leq2K|v_{\rm in}| .
\end{equation}
Writing \(\vartheta=\arcsin\sqrt{\langle\mathcal P\rangle}\) for the leakage angle gives \(|\dot\vartheta|\leq\|\mathcal PG\mathcal Q\|\). The localized input has \(\vartheta(0)=0\), while fidelity \(F\) with the Dicke state requires \(\vartheta\geq[\arcsin\sqrt{K/N}-\arccos\sqrt F]_+\). Integration gives Eq.~\eqref{eq:endpoint-cut-bound}.

\section{Microscopic validation}
\label{app:eff-micro}
The effective value curves are optimized first for \(\sect{6}{3}\), \(\sect{8}{4}\), and \(\sect{7}{3}\) at \(q=\delta_{\rm in}/\delta_{\rm sh}\in\{0.5,1,2\}\). The microscopic runs then use \(\gamsh=\pi/2-\delta_{\rm sh}\) and \(\gamin=\pi/2-q\delta_{\rm sh}\) with the exact chronological partial-SWAP word. The two reported component detunings are \(\delta_{\rm sh}=0.0125\) and \(0.00625\). On each fixed ray, \(\varepsilon=\delta_{\rm sh}\sqrt{1+q^2}\) and \(\tau=\tau_{\rm sh}\sqrt{1+q^2}\) with \(\tau_{\rm sh}=\delta_{\rm sh}T\). Table~\ref{tab:eff-micro-agreement} reports the relative contour discrepancy at both detunings,
\begin{equation}
 \Delta(F)=100\!\left[
 \frac{\delta_{\rm sh}\hat T_{\min}^{\rm micro}(F)}
 {\hat\tau_{\min}^{\rm eff}(F)}-1\right]\!\% .
 \label{eq:eff-micro-delta}
\end{equation}
The radial conversion multiplies numerator and denominator by the same factor \(\sqrt{1+q^2}\), so Eq.~\eqref{eq:eff-micro-delta} is unchanged in the radial convention of Eq.~\eqref{eq:radial-detuning}.

\begin{table}[!tbp]
\caption{Effective-to-microscopic hitting-time comparison at the two shuttle-component detunings \(\delta_{\rm sh}=0.0125\) and \(0.00625\). Entries are the relative discrepancies \(\Delta(F)\) of Eq.~\eqref{eq:eff-micro-delta}; positive values mean that the microscopic crossing occurs at a larger rescaled time than the independently optimized effective crossing. The last column is the median absolute discrepancy across the four fidelities.}
\label{tab:eff-micro-agreement}
\centering
\scriptsize
\setlength{\tabcolsep}{2.2pt}
\begin{tabular}{lclrrrrr}
\toprule
Sector & $q$ & $\delta_{\rm sh}$ & $.95$ & $.98$ & $.99$ & $.999$ & Med. $|\Delta|$ \\
\midrule
\sect{6}{3} & 0.5 & 0.0125 & +0.9\% & $-$5.7\% & $-$4.8\% & $-$1.5\% & 3.1\% \\
 &  & 0.00625 & +2.0\% & $-$2.2\% & $-$4.0\% & $-$0.8\% & 2.1\% \\
\addlinespace[1pt]
\sect{6}{3} & 1.0 & 0.0125 & +0.1\% & +0.7\% & $-$0.3\% & $-$4.5\% & 0.5\% \\
 &  & 0.00625 & +1.3\% & +1.8\% & +0.7\% & $-$2.1\% & 1.5\% \\
\addlinespace[1pt]
\sect{6}{3} & 2.0 & 0.0125 & +0.1\% & $-$1.8\% & $-$0.6\% & +0.6\% & 0.6\% \\
 &  & 0.00625 & +1.6\% & $-$0.5\% & +3.2\% & +2.8\% & 2.2\% \\
\addlinespace[1pt]
\sect{8}{4} & 0.5 & 0.0125 & $-$0.8\% & $-$1.0\% & $-$5.3\% & $-$2.9\% & 1.9\% \\
 &  & 0.00625 & +0.3\% & +2.1\% & $-$2.7\% & +0.3\% & 1.2\% \\
\addlinespace[1pt]
\sect{8}{4} & 1.0 & 0.0125 & +2.4\% & $-$1.7\% & $-$1.1\% & $-$2.0\% & 1.9\% \\
 &  & 0.00625 & +0.8\% & $-$3.0\% & +0.1\% & $-$2.0\% & 1.4\% \\
\addlinespace[1pt]
\sect{8}{4} & 2.0 & 0.0125 & $-$6.4\% & $-$5.0\% & $-$3.1\% & $-$12.9\% & 5.7\% \\
 &  & 0.00625 & $-$6.4\% & $-$7.1\% & $-$3.1\% & $-$4.9\% & 5.6\% \\
\addlinespace[1pt]
\sect{7}{3} & 0.5 & 0.0125 & $-$0.5\% & +2.4\% & $-$3.2\% & $-$3.7\% & 2.8\% \\
 &  & 0.00625 & +1.9\% & +3.5\% & $-$0.4\% & $-$2.1\% & 2.0\% \\
\addlinespace[1pt]
\sect{7}{3} & 1.0 & 0.0125 & +0.9\% & $-$2.7\% & $-$2.3\% & $-$5.2\% & 2.5\% \\
 &  & 0.00625 & +2.4\% & $-$1.4\% & +0.3\% & $-$3.0\% & 1.9\% \\
\addlinespace[1pt]
\sect{7}{3} & 2.0 & 0.0125 & $-$2.5\% & $-$2.0\% & $-$1.3\% & $-$7.9\% & 2.3\% \\
 &  & 0.00625 & $-$4.8\% & $-$2.0\% & $-$3.2\% & $-$4.7\% & 4.0\% \\
\bottomrule
\end{tabular}
\end{table}

All 36 sector--direction--fidelity cells remain within \(10\%\) at the smaller detuning, and most discrepancies are a few percent. Halving the detuning from \(0.0125\) to \(0.00625\) reduces the largest discrepancy from \(12.9\%\), in the single cell \(\sect{8}{4}\), \(q=2\), \(F=0.999\), to \(7.1\%\), while the median absolute discrepancy stays near \(2\%\) at both detunings and the absolute discrepancy decreases in 17 of the 36 cells. Because the horizon is an integer, one round corresponds to \(1\)--\(5\%\) of \(\hat T_{\min}\) at \(\delta_{\rm sh}=0.0125\) and to half of that at \(0.00625\), so the convergence is resolved in the tail of the distribution and the bulk is already at the resolution of the search.

\section{Target geometry}
\label{app:target-geometry}
For nonnegative exchange weights, \(\mathcal L_{\hat{\bm v}}\) is positive semidefinite. If the exchange graph is connected, its induced graph on \(K\)-subsets is connected for \(0<K<N\), and its one-dimensional null space is spanned by the Dicke state. Let \(\lambda_2>0\) and \(\lambda_{\max}\) be the smallest positive and the largest eigenvalue. With \(\Phi\) from Eq.~\eqref{eq:phase-terminal-payoff},
\begin{equation}
 \lambda_2[1-\Phi(\psi)]
 \leq E_G(\psi)
 \leq\lambda_{\max}[1-\Phi(\psi)].
 \label{eq:gauge-frustration-bounds}
\end{equation}
These inequalities follow by decomposing \(D(\Theta)\ket{\psi}\) into its Dicke component and its orthogonal complement, then minimizing over the phase torus. Zero frustration requires equal magnitudes with phases realizable by the available local rotations.

For a localized \(K\)-subset \(s\), let \(W_s\) be the total weight of exchange edges with one occupied endpoint and one empty endpoint. If the target \(\ket{\psi_{\rm tar}}\) has equal magnitudes \(1/\sqrt C\), its initial fidelity derivative under \(H_\Theta=D(\Theta)^\dagger H_{\hat{\bm v}}D(\Theta)\) obeys
\begin{equation}
\begin{aligned}
 \dot F_\Theta(s)
 &=2\operatorname{Re}\!\left[
 \langle s|\psi_{\rm tar}\rangle
 \langle\psi_{\rm tar}|(-iH_\Theta)|s\rangle\right],\\
 \dot F_\Theta(s)&\leq\frac{2W_s}{C}.
\end{aligned}
\label{eq:steering-bound}
\end{equation}
For the Dicke target, choose phase \(\pi/2\) on occupied sites and zero on empty sites. Every outgoing exchange then contributes with the same sign, and the bound is attained. A gauge shift gives the same result for local-phase Dicke targets. General equal-magnitude targets impose simultaneous site-phase conditions through their configuration-phase differences, and incompatible conditions produce a positive steering deficit.

The geometry study uses the same positive diagonal defect graph in \(\sect{9}{4}\) and \(\sect{10}{5}\). Direct differentiation of the collision map reproduces the graph weights to relative error below \(1.1\times10^{-10}\). Dicke controls are gauge-flat, while the phase searches find positive frustration and a steering-deficit witness for every sampled phase-uniform and random-sign target. Repeating the phase optimization with 32 starts gives the same classifications.

\begin{table}[!tbp]
\caption{Median paired difference \(\Delta\hat T_{\min}=\hat T_{\min}(1.5,1.5)-\hat T_{\min}(1,1)\), in rounds, for the 16-target matched study. Positive values mean that the coupling \((1.5,1.5)\), at \(\varepsilon=0.10\) from full SWAP, is slower.}
\label{tab:target-deltas}
\begin{ruledtabular}
\begin{tabular}{lrrrr}
Sector / family & $.95$ & $.98$ & $.99$ & $.999$\\
\colrule
\sect{9}{4} phase-uniform & 6 & 8 & 8 & 11\\
\sect{9}{4} random-sign & 6 & 7 & 8 & 10\\
\sect{9}{4} Haar & 6 & 7.5 & 8 & 10.5\\
\sect{9}{4} random-positive & 0 & 3 & 5.5 & 11\\
\sect{10}{5} phase-uniform & 9 & 11 & 12 & 15\\
\sect{10}{5} random-sign & 9 & 10 & 12 & 15\\
\sect{10}{5} Haar & 9 & 11 & 12 & 16\\
\sect{10}{5} random-positive & 0 & 8 & 13 & 16\\
\end{tabular}
\end{ruledtabular}
\end{table}

For the phase-uniform, random-sign, and Haar targets, every one of the 16 paired differences is positive at every reported fidelity in both systems. Random-positive targets are mixed at \(F=0.95\) and uniformly positive at the three higher levels. The Dicke target has the opposite sign.

\bibliography{SD_refs}

\end{document}